\documentclass[aps,prb,reprint,superscriptaddress,nobibnotes]{revtex4-1}
\usepackage{graphicx,epsfig}
\usepackage{amsmath,amssymb,bm}
\usepackage{color}
\usepackage[usenames,dvipsnames]{xcolor}
\usepackage[colorlinks=true,linkcolor=Maroon,citecolor=OliveGreen,urlcolor=Blue,linktoc=page]{hyperref}
\usepackage{placeins}
\usepackage{ulem}
\usepackage[symbol]{footmisc} %
\usepackage{multirow}

\begin{document}

%=============================================================================
%=============================================================================

\title{Growth and characterization of large (Y,La)TiO$_3$ and (Y,Ca)TiO$_3$ single crystals}    
\author{S. Hameed}
\thanks{Corresponding authors: hamee007@umn.edu, greven@umn.edu}
\affiliation{School of Physics and Astronomy, University of Minnesota, Minneapolis, MN 55455, U.S.A}
\author{J. Joe}
\affiliation{School of Physics and Astronomy, University of Minnesota, Minneapolis, MN 55455, U.S.A}
\author{L. R. Thoutam}
\affiliation{Department of Chemical Engineering and Materials Science, University of Minnesota, Minneapolis, MN 55455, U.S.A}
\affiliation{Department of Electronics and Communications Engineering, SR University, Warangal Urban, Telangana 506371, India}
\author{J. Garcia-Barriocanal}
\affiliation{Characterization Facility, University of Minnesota, Minneapolis, MN 55455, U.S.A}
\author{B. Yu}
\affiliation{School of Physics and Astronomy, University of Minnesota, Minneapolis, MN 55455, U.S.A}
\author{G. Yu}
\affiliation{Characterization Facility, University of Minnesota, Minneapolis, MN 55455, U.S.A}
\affiliation{Informatics Institute, University of Minnesota, Minneapolis, Minnesota 55455, USA}
\author{S. Chi}
\affiliation{Neutron Scattering Division, Oak Ridge National Laboratory, Oak Ridge, Tennessee 37831, USA}
\author{T. Hong}
\affiliation{Neutron Scattering Division, Oak Ridge National Laboratory, Oak Ridge, Tennessee 37831, USA}
\author{T. J. Williams}
\affiliation{Neutron Scattering Division, Oak Ridge National Laboratory, Oak Ridge, Tennessee 37831, USA}
\author{J. W. Freeland}
\affiliation{X-ray Science Division, Argonne National Laboratory, Argonne, IL 60439, USA}
\author{P. M. Gehring}
\affiliation{NIST Center for Neutron Research, National Institute of Standards and Technology, Gaithersburg, Maryland 20899, USA}
\author{Z. Xu}
\affiliation{NIST Center for Neutron Research, National Institute of Standards and Technology, Gaithersburg, Maryland 20899, USA}
\affiliation{Department of Materials Science and Engineering, University of Maryland, College Park, Maryland 20742, USA}
\author{M. Matsuda}
\affiliation{Neutron Scattering Division, Oak Ridge National Laboratory, Oak Ridge, Tennessee 37831, USA}
\author{B. Jalan}
\affiliation{Department of Chemical Engineering and Materials Science, University of Minnesota, Minneapolis, MN 55455, U.S.A}
\author{M. Greven}
\thanks{Corresponding authors: hamee007@umn.edu, greven@umn.edu}
\affiliation{School of Physics and Astronomy, University of Minnesota, Minneapolis, MN 55455, U.S.A}

%=============================================================================
%=============================================================================

\widetext
\date{\today}

\begin{abstract}
The Mott-insulating rare-earth titanates (RTiO$_3$, R being a rare-earth ion) are an important class of materials that encompasses interesting spin-orbital phases as well as ferromagnet-antiferromagnet and insulator-metal transitions. The growth of these materials has been plagued by difficulties related to overoxidation, which arises from a strong tendency of Ti$^{3+}$ to oxidize to Ti$^{4+}$. We describe our efforts to grow sizable single crystals of YTiO$_3$ and its La-substituted and Ca-doped variants with the optical travelling-solvent floating-zone technique. We present sample characterization {\it via} chemical composition analysis, magnetometry, charge transport, neutron scattering, x-ray absorption spectroscopy and x-ray magnetic circular dichroism to understand macroscopic physical property variations associated with overoxidation. 
Furthermore, we demonstrate a good signal-to-noise ratio in inelastic magnetic neutron scattering measurements of spin-wave excitations.
A superconducting impurity phase, found to appear in Ca-doped samples at high doping levels, is identified as TiO.

% check carefully that we are not generalizing to RTiO3. The findings might in fact be specific to Y(La,Ca)TO as seen in Roth thesis
% need to add statements about picking samples for neutron scattering
% check Fig 7 caption
% Diffuse data on YLTO barely shows an impurity phase
\end{abstract}
\maketitle

%\clearpage

%-------------------------------------------------------------------
%Introduction
%-------------------------------------------------------------------

\section{Introduction}

The Mott-insulating rare-earth titanates RTiO$_3$ exhibit myriad magnetic and electronic phases as a result of a strong coupling among spin, orbital, charge and lattice degrees of freedom \cite{Mochizuki2004}. These oxides have a GdFeO$_3$-type distorted perovskite structure, with tilts and rotations of the oxygen octahedra that arise from the size mismatch between the R and Ti ions. The distortion, and the associated deviation of the Ti-O-Ti bond angle from 180$^\text{o}$, increases with decreasing R-ion radius \cite{MacLean1979} and has a strong influence on the superexchange involving the Ti $3d$ $t_{2g}^1$ spin-1/2 moments. 
%MG: superexchange (as opposed to direct exchange) involves intervening anions
This leads to a rare antiferromagnetic-ferromagnetic phase transition/cross-over; materials with a large R-ion radius (La, Sm, Nd, {\it etc.}) display an antiferromagnetic ground state, whereas those with a small R-ion radius (Y, Gd, Dy, {\it etc.}) display a ferromagnetic ground state \cite{Greedan1985,Zhou2005a,Mochizuki2004}. The phase-transition/cross-over can also be induced in solid-solution systems formed by isovalent substitution, $e.g.$, Y$_{1-x}$La$_x$TiO$_3$ \cite{Goral1982,Okimoto1995,Zhou2005,HameedYLa2021} and Sm$_{1-x}$Gd$_x$TiO$_3$ \cite{Amow2000}. 

It is also possible to tune RTiO$_3$ from a Mott insulator to a metallic state $via$ hole doping, which can be achieved by substituting the trivalent R$^{3+}$ ion with a divalent A$^{2+}$ ion such as Ca$^{2+}$ or Sr$^{2+}$ to form R$_{1-y}$A$_y$TiO$_3$. Intriguingly, the critical doping $y = y_c$ up to which the Mott-insulating state survives can be located significantly away from half-filling ($y=0$), with $y_c$ ranging from 0.05 in La$_{1-y}$Sr$_y$TiO$_3$ \cite{Tokura1993La} to 0.35 in Y$_{1-y}$Ca$_y$TiO$_3$ \cite{Tokura1993,HameedYCa2021}. Recent work on Y$_{1-y}$Ca$_y$TiO$_3$ showed that this insulator-metal transition involves electronic phase separation  \cite{HameedYCa2021}.

Recent years have also witnessed a surge of interest in RTiO$_3$-based heterostructures, in a search for novel electronic properties and correlated electronic states at interfaces \cite{Xu2016a,Xu2016b,Cao2016a,Raghavan2015,Moetakef2012,Grisolia2016}. For instance, a ferromagnetic-like state was induced in a rare-earth nickelate (RNiO$_3$) by forming an interface with GdTiO$_3$ \cite{Grisolia2016}. Another example is the control of charge carrier densities in NdTiO$_3$/SrTiO$_3$ heterostructures \cite{Xu2016a,Xu2016b}. It has thus become increasingly important to fully characterize and understand the properties of the bulk compounds. One of the primary issues that has plagued the study of the electronic and magnetic properties of RTiO$_3$ is a strong tendency of Ti$^{3+}$ to oxidize to Ti$^{4+}$, which makes it extremely difficult to obtain stoichiometric samples. Whereas this overoxidation effect can result from both cation vacancies and oxygen interstitials, the latter has been shown to be energetically favored \cite{Xu2016b}. The overoxidation effect has been the subject of recent studies, particularly of films, and has been shown to significantly affect magnetic and electronic properties \cite{Aeschlimann2018,Aeschlimann2021}.

Here we describe our efforts to grow large single crystals of La-substituted and Ca-doped YTiO$_3$ with the traveling solvent floating zone (TSFZ) technique, to enable neutron scattering and muon spin rotation studies of the magnetic ground state \cite{HameedYLa2021}. We carefully characterized the changes in properties that can appear across large single crystals using chemical composition analysis, magnetometry, charge transport, elastic and inelastic neutron scattering, x-ray diffraction, x-ray absorption spectroscopy and x-ray magnetic circular dichroism. We discuss the difficulties associated with obtaining homogeneous large single crystals, and the procedure that we adopted to prepare sizable samples suitable for inelastic neutron scattering experiments \cite{HameedSW2021}. We also identify and characterize a superconducting impurity phase of TiO that appears at large Ca-doping levels in Y$_{1-y}$Ca$_y$TiO$_3$. Our goal here is to provide a comprehensive report on the growth of large single crystals of Y$_{1-x}$La$_x$TiO$_3$ and Y$_{1-y}$Ca$_y$TiO$_3$, so as to aid future studies of this important class of quantum materials. 
Some of these crystals were used in recent studies of the ferromagnet-antiferromagnet transition in Y$_{1-x}$La$_x$TiO$_3$ \cite{HameedYLa2021}, the Mott insulator-metal transition in Y$_{1-y}$Ca$_y$TiO$_3$ \cite{HameedYCa2021}, and the effect of uniaxial strain on the magnetic ground state in these systems \cite{Ana2021}.

This paper is organized as follows: Section~\ref{section:expmeth} describes the single-crystal growth and experimental details. Sections~\ref{section:YTO}, \ref{section:YLTO} and \ref{section:YCTO} describe the characterization of single crystals of YTiO$_3$, Y$_{1-x}$La$_x$TiO$_3$ and Y$_{1-y}$Ca$_y$TiO$_3$, respectively. Section~\ref{section:Neutron} discusses sample characterization $via$ elastic and inelastic neutron scattering. Finally, we summarize our main conclusions in Section~\ref{section:conc}.

%-------------------------------------------------------------------
%Experimental Methods
%-------------------------------------------------------------------

\section{Experimental methods}
\label{section:expmeth}

Single crystals of Y$_{1-x}$La$_{x}$TiO$_{3}$ and Y$_{1-y}$Ca$_{y}$TiO$_{3}$ were melt-grown with the optical TSFZ technique. The starting materials were La$_{2}$O$_{3}$, Y$_{2}$O$_{3}$ and Ti$_{2}$O$_{3}$ for Y$_{1-x}$La$_{x}$TiO$_{3}$, and Y$_{2}$O$_{3}$, CaTiO$_3$ and Ti$_{2}$O$_{3}$ for Y$_{1-y}$Ca$_{y}$TiO$_{3}$. In order to prevent the oxidation of Ti$^{3+}$ to Ti$^{4+}$, we {\it started} with an oxygen-deficient composition Y$_{1-x}$La(Ca)$_{x}$TiO$_{3-\delta}$, similar to prior work \cite{Goral1982,Okimoto1995}. This was achieved by adding Ti powder to the starting materials. The idea here is to compensate for residual oxygen in the furnace that could create oxygen interstitials in the grown crystal \cite{Xu2016b}. We emphasize that $\delta$ simply refers to the composition of the {\it starting} materials in the synthesis; this does not imply that the grown crystals have the same value of $\delta$. The La$_{2}$O$_{3}$, CaTiO$_3$ and Y$_{2}$O$_{3}$ powders were pre-dried in air at 1000$^{\text{o}}$C for 12 hours. The starting materials were then mixed in the stoichiometric ratio Y$_{1-x}$La(Ca)$_{x}$TiO$_{3-\delta}$, and subsequently pressed at 70 MPa into two rods with a diameter of 6 mm and lengths of 20 mm and 100 mm for use as seed and feed rods, respectively. The feed and seed rods were then loaded into a Crystal Systems, Inc. four-mirror optical TSFZ furnace. The growths were performed at rates of 2 to 5 mm/h in a reducing atmosphere with a mixed gas of 5\%H$_{2}$/95\%Ar at a pressure of 5 bar in the case of the La-substituted system, and pure Ar gas at a pressure of 5 bar in the case of the Ca-doped system. The growth chamber was flushed with the gas before the start of each growth. In the case of YTiO$_3$, a pyrochlore Y$_2$Ti$_2$O$_7$ impurity phase 
%MG: this is still not quite clear. What does 'predominantly' mean in this context? What was found in the case of the Ca/La systems? - we never did the growth without flushing for Ca/La systems
was observed to form in the absence of this flushing step due to the large amount of residual oxygen in the chamber. The choice of $\delta$ for different compositions is detailed below, in the respective Sections.

Laue diffraction measurements confirmed the single-crystal nature of the samples and enabled the determination of the crystallographic axes. YTiO$_3$ and its La-substituted and Ca-doped variants have a $Pbnm$ orthorhombic structure at room temperature. We use the orthorhombic notation throughout, indicated by the subscript ``o''.
% MG: I think you need to specify the structural symmetry here, specify typical room temperature lattice constants, and define a as [100], etc.
Chemical composition analysis was performed $via$ wavelength-dispersive spectroscopy (WDS). Note that the oxygen off-stoichiometry cannot be obtained from WDS and was not determined for our samples. Magnetic-susceptibility measurements were performed with a Quantum Design, Inc. Magnetic Property Measurement System (MPMS). For DC susceptibility measurements, we used the dc SQUID magnetometer, whereas AC susceptibility measurements were performed with a home-built probe that consists of two matched detection coils, with the excitation field provided by the built-in MPMS coil. The magnetic field was applied along the sample's $c$-axis. Resistivity measurements for YTiO$_3$ were carried out with a Quantum Design, Inc. PPMS Dynacool system. A Keithley 2612B sourcemeter was used in the four-wire configuration to source the current and measure the voltage. The samples were cut into a square shape with an $ab$-plane surface, the $a$- and $b$-axes being the sides of the square. The sample surfaces were polished with lapping films to 1 $\mu$m grade, and electrical contacts were made with gold wires and silver paint. All samples were measured in a van der Pauw geometry. X-ray absorption spectroscopy/x-ray magnetic circular dichroism (XAS/XMCD) measurements were carried out at beam-line 4-ID-C of the Advanced Photon Source at Argonne National Laboratory. The measurements reported here were obtained in total fluorescence yield (TFY) detection mode. A reference SrTiO$_3$ sample was used to align the spectra energy scales to within 0.1 eV. The surfaces of the samples used for XAS/XMCD measurements were cut parallel to the [001] plane and polished to a roughness of $\sim$ 0.3 $\mu$m using polycrystalline diamond suspension. The x-rays were incident at an angle of 30$^{\text{o}}$ with respect to the
$c$-axis. A 5-T magnetic field was applied along the x-ray beam for the XMCD measurements. X-ray diffraction (XRD) measurements were performed on a Bruker D8 Discover x-ray diffractometer, equipped with a Co K$\alpha$ x-ray source and a VANTE-500 area detector.

Triple-axis neutron scattering experiments were performed with the HB-1 and HB-3 thermal triple-axis spectrometers and the CG-4C cold-triple axis spectrometer at the High Flux Isotope Reactor, Oak Ridge National Laboratory, and with the BT-7 thermal triple-axis spectrometer at the NIST Center for Neutron research. The final energies of the scattered neutrons were fixed at 14.7 meV for the thermal triple-axis measurements and 4.5 meV for the cold triple-axis measurements. We used horizontal beam collimations of $48'-80'-$sample$-80'-120'$ at HB-1 and HB-3, open$-80'-$sample$-80'-120'$ at BT-7, and open$-$open$-$sample$-80'-$open at CG-4C. In order to eliminate higher-order neutrons, PG filters were used at HB-1, HB-3 and BT-7, and a cooled Be filter was used at CG-4C. A $^4$He-cryostat was used to reach temperatures down to 1.5 K. The samples were mounted in the (0$KL$) scattering plane.

%-------------------------------------------------------------------
%YTO characterization
%-------------------------------------------------------------------

\section{Results}

\subsection{Characterization of YTiO$_3$ single crystals}
\label{section:YTO}

\begin{figure}
\includegraphics[width=0.3\textwidth]{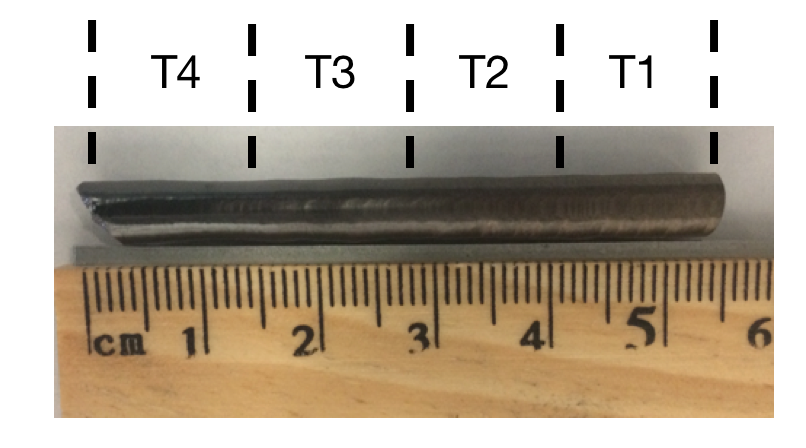}
\caption{Picture of a large YTiO$_3$ single crystal grown with the TSFZ method. The sectioning of the single crystal used for characterization is indicated.}
\label{fig:YTOSC}
\end{figure}
%MG: Your notation uses parentheses in this figure, but not in the text. I would omit them. - done

\begin{figure}
\includegraphics[width=0.4\textwidth]{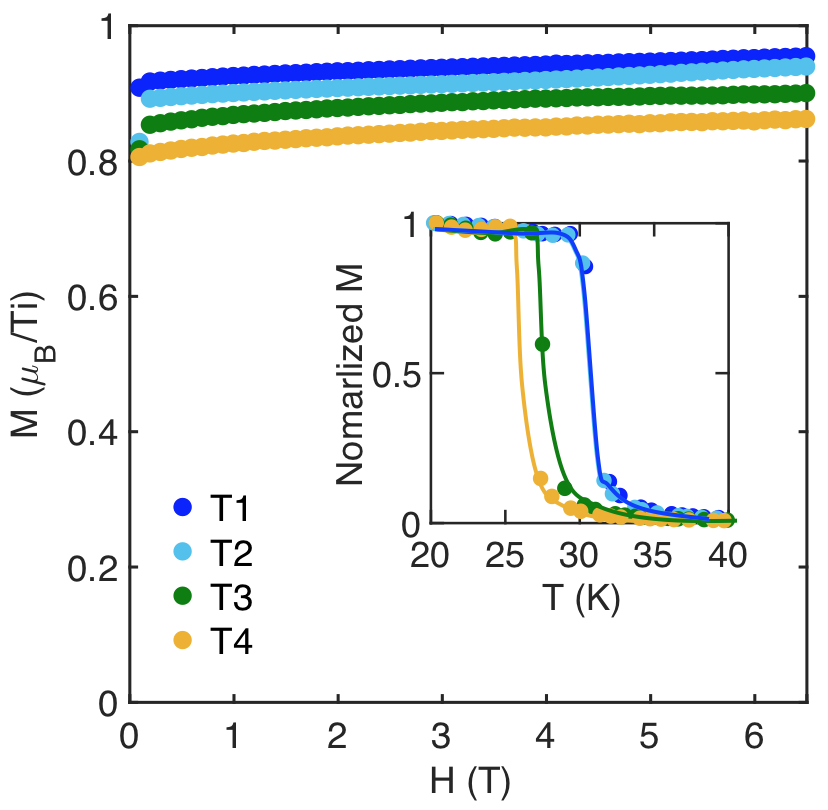}
\caption{Magnetic-field dependence of the magnetic moment per Ti ion for the YTiO$_3$ samples, obtained along the $c$-axis at $T = 5$ K. The samples were zero-field cooled and measured in an increasing field. Inset: Temperature dependence of the field-cooled $c$-axis magnetization, obtained in an applied magnetic field of 50 Oe. The magnetization values are normalized to those at 20 K.}
\label{fig:YTOMag}
\end{figure}

\begin{table}[h!]
  \begin{center}
    \caption{Measured chemical composition, Curie temperature, magnetic moment at 5 K and 6.5 T, transport activation energy, and room-temperature resistivity of the YTiO$_{3}$ single-crystal samples.}
    \label{tab:table1}
    \vspace*{5mm}

    \begin{tabular}{c|c|c|c|c|c}
    \textbf{Sample} & \textbf{Y:Ti} & \textbf{$T_C$ } & \textbf{FM} & \textbf{$E_a$ } & \textbf{$\rho$} \\
    & & \textbf{(K)} & \textbf{moment} & \textbf{(meV)} & \textbf{(300 K)} \\
    \textbf{} & \textbf{} & \textbf{} & \textbf{($\mu$$_{\text{B}}$/Ti)} & \textbf{} & \textbf{($\Omega$cm)} \\
      \hline
      T1 & 0.964(7) & 30.8(4) & 0.96(2) & 316.8(8) & 260(5)\\
      T2 & 0.978(8) & 30.8(4) & 0.94(1) & 295.4(8) & 112(3)\\
      T3 & 0.965(9) & 27.5(5) & 0.90(1) & 219.9(4) & 11.2(2)\\
      T4 & 0.944(4) & 26.5(5) & 0.86(1) & 221.3(9) & 15.5(5)\\
      
    \end{tabular}
  \end{center}
\end{table}

As detailed in Section~\ref{section:expmeth}, we used an oxygen-deficient \textit{starting} composition to grow the single crystals. For the case of YTiO$_3$, several different values of $\delta$ were attempted in the range $0 - 0.08$. We found that $\delta$ = 0.04 yielded the highest Curie temperature of $T_C$ $\sim$ 30 K, as determined by magnetometry. For this step of optimizing $\delta$ in the growth process, we characterized a single crystal piece picked from the end of the growth. However, as detailed below, we found that significant inhomogeneities exist in the magnetic and transport properties across the length of the growth.

% MuSR yielding 100% magnetic volume fraction tells us that delta is good. But this is questionable for 20% and 30% samples I guess

Figure~\ref{fig:YTOSC} shows an example of a large YTiO$_3$ single crystal. The characterization was performed by cutting the crystal into four approximately equal-length parts, as indicated in the figure, and by selecting a sample from the central region of each cut-out portion. 
The naming convention here is such that the numbered suffix represents the position of the sample from the end of the growth. We characterized the samples using magnetometry, charge transport, chemical composition analysis and XAS/XMCD. 

\begin{figure}
\includegraphics[width=0.4\textwidth]{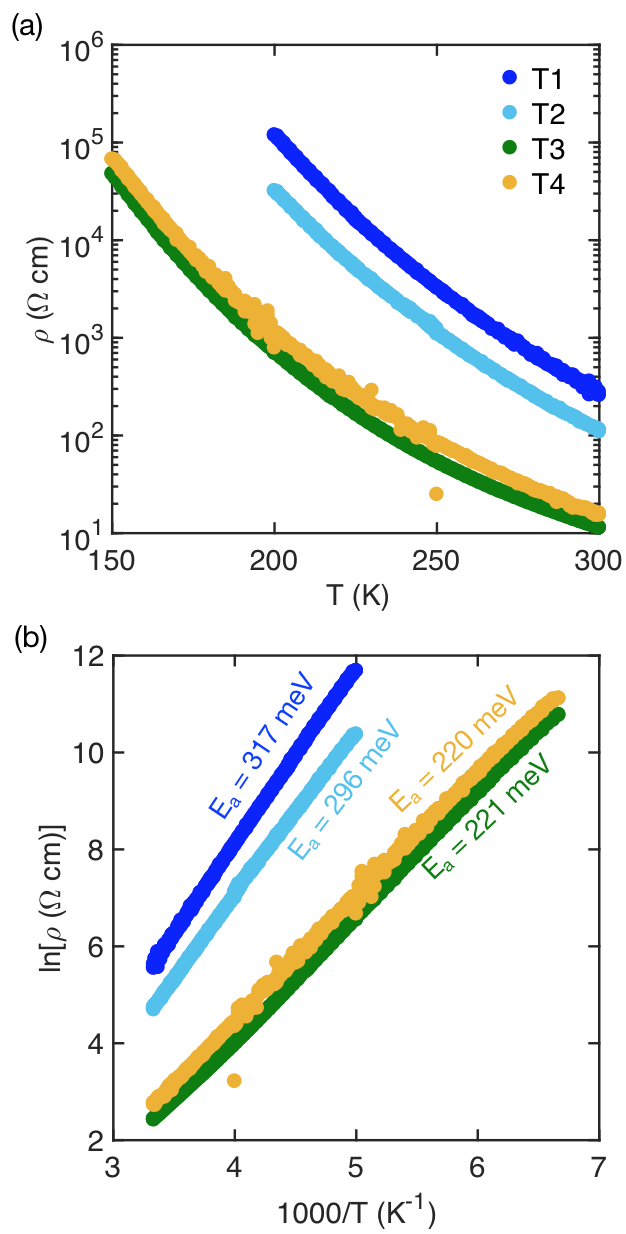}
\caption{(a) Temperature dependence of resistivity for the YTiO$_3$ samples. (b) Arrhenius plot for the resistivities in (a), along with the corresponding activation energies.}
\label{fig:YTO_transport}
\end{figure}
%MG: figure caption was incorrect - please add

Figure~\ref{fig:YTOMag} shows the magnetic-field dependence of the magnetization $M$ obtained at 5 K for the four YTiO$_3$ samples. The values of $M$ obtained at 6.5 T are summarized in Table~\ref{tab:table1}. Whereas samples T1 and T2 are seen to exhibit closely similar $M(H)$ behavior, a clear decrease in $M$ is observed for sample T3, and an even further decrease is observed for T4. The inset to Fig.~\ref{fig:YTOMag} displays the temperature dependence of the field-cooled magnetization obtained in an applied magnetic field of 50 Oe. The Curie temperature is defined at the midpoint of the transition observed in the $M$ $vs.$ $T$ curves, and summarized in Table~\ref{tab:table1}. Samples T1 and T2 exhibit closely similar Curie temperatures, whereas a decrease in $T_C$ is observed for sample T3, and a further decrease for sample T4.

As a next characterization step, we measured the temperature dependence of the resistivity, as shown in Fig.~\ref{fig:YTO_transport}. 
%MG: there is a glitch with the figure numbering
All samples display insulating behavior. Further, the temperature dependence of the resistivity is well fit to thermally-activated behavior, $\rho \propto e^{-E_a/k_BT}$, in agreement with prior reports \cite{Taguchi1993,Zhou2005,Zhou2005a,Yue2020}. The mechanism for the activated charge transport was recently shown to be associated with small hole-polaron hopping \cite{Yue2020}. The room temperature resistivity and the fit results for the activation energy $E_a$ are displayed in Table~\ref{tab:table1}. The room-temperature resistivity of samples T3 and T4 are observed to be an order of magnitude lower than those of samples T1 and T2. Furthermore, samples T3 and T4 exhibit a smaller activation energy.

In order to understand the origin of the different magnetic and transport properties obtained for the YTiO$_3$ samples, we carried out detailed chemical composition analysis using WDS. The results are summarized in Table~\ref{tab:table1}. Note that the uncertainties indicated for the Y:Ti ratio are standard uncertainties determined by measuring the composition at five different spots on the sample. The Y:Ti ratio for the samples T1, T2 and T3 are observed to be similar, within error, whereas sample T4 has a slightly lower Y:Ti ratio. However, as seen from transport and magnetometry measurements, a clear difference exists between the properties of sample T3 and samples T1 and T2. Therefore, the observed differences cannot be entirely due to a slightly different Y:Ti ratio, which points to the possibility of oxygen interstitials, the formation of which was previously shown to be energetically favorable in the RTiO$_3$ class of materials \cite{Xu2016a}. Unfortunately, this information is not accessible $via$ WDS. Since both cation vacancies and oxygen interstitials would lead to a conversion from Ti$^{3+}$ to Ti$^{4+}$, we attempted to characterize the extent of such overoxidation using XAS, as described below.

\begin{figure}
\includegraphics[width=0.45\textwidth]{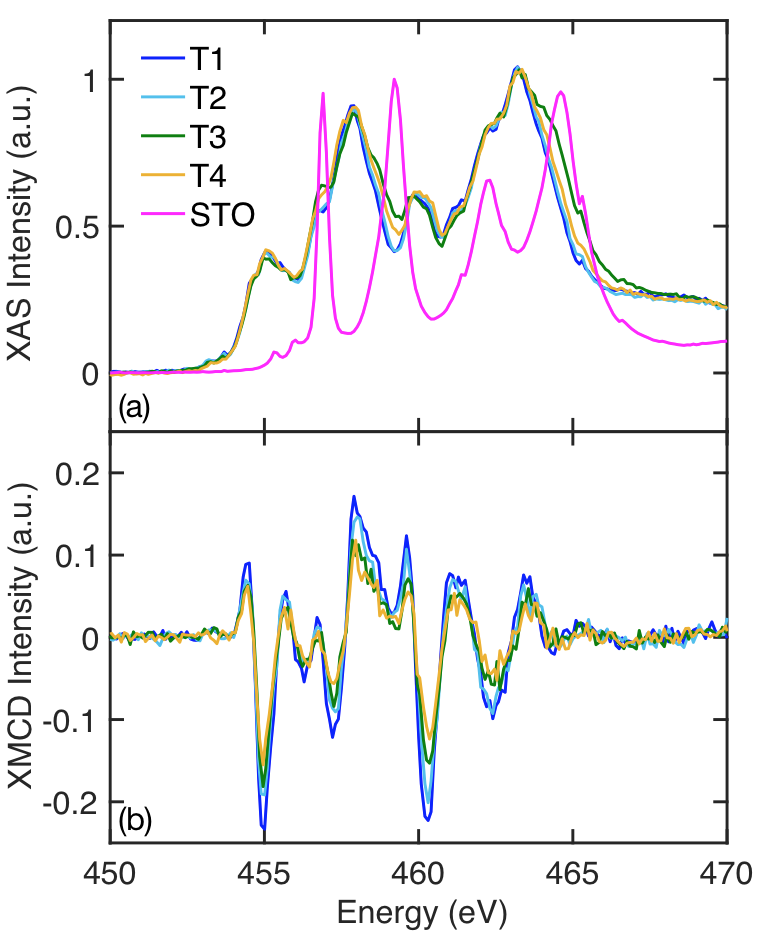}
\caption{(a) Ti $L_{2,3}$-edge XAS results for four YTiO$_3$ samples. The data measured at the lowest energy were subtracted, and the net intensity was subsequently normalized at the highest peak. A Ti$^{4+}$ reference spectrum obtained from a SrTiO$_3$ (STO) sample is displayed as well. (b) XMCD intensities at the Ti $L_{2,3}$-edge at 6 K, in the ferromagnetic phase. The data were normalized by the intensity of the largest Ti $L_{2,3}$-edge XAS peak.}
\label{fig:XAS}
\end{figure}

XAS is a powerful technique capable of distinguishing different valence states of elements. Figure~\ref{fig:XAS}(a) shows the XAS spectra obtained at the Ti $L_{2,3}$-edge. Clearly, the XAS spectra for samples T1 and T2 approximately coincide, whereas samples T3 and T4 display comparatively broadened peaks. A comparison with reference Ti$^{4+}$ spectra obtained for a SrTiO$_3$ single crystal clearly indicates that the additional XAS intensity in samples T3 and T4 appears in the energy range associated with the Ti$^{4+}$ spectral peaks, indicative of an overoxidation in samples T3 and T4. 
%MG: Is the term 'overoxidation' used in the literature? - SH: yes, the papers by Manuel Bibes that I cite in the intro uses this term
Such a partial conversion of Ti$^{3+}$ to Ti$^{4+}$ has been reported in several other XAS works on the RTiO$_3$ class of materials \cite{Ulrich2008,Aeschlimann2018,Aeschlimann2021}. In order to confirm the effect of such oxidation on the magnetic properties, we performed XMCD measurements at 6 K, deep in the ferromagnetic phase. As seen in Fig.~\ref{fig:XAS}(b), a gradual reduction in the XMCD signal is observed from sample T1 to sample T4, indicative of a weakening of the magnetic order, in agreement with the magnetometry results.

Our results clearly illustrate that overoxidation $via$ formation of cation-vacancies or oxygen interstitials can strongly affect the magnetic and transport properties of YTiO$_3$. It must be noted, however, that there could be additional causes of these differences, such as extrinsic inhomogeneous strains that may nucleate across a large single crystal during growth. Ideally, the strain would be removed by annealing. However, the extreme tendency of Ti$^{3+}$ to convert to Ti$^{4+}$ in the presence of traces of oxygen renders such annealing processes extremely difficult for these materials.

%-------------------------------------------------------------------
%YLTO and YCTO characterization
%-------------------------------------------------------------------

\subsection{Characterization of Y$_{1-x}$La$_{x}$TiO$_{3}$ single crystals}
\label{section:YLTO}

\begin{figure}
\includegraphics[width=0.4\textwidth]{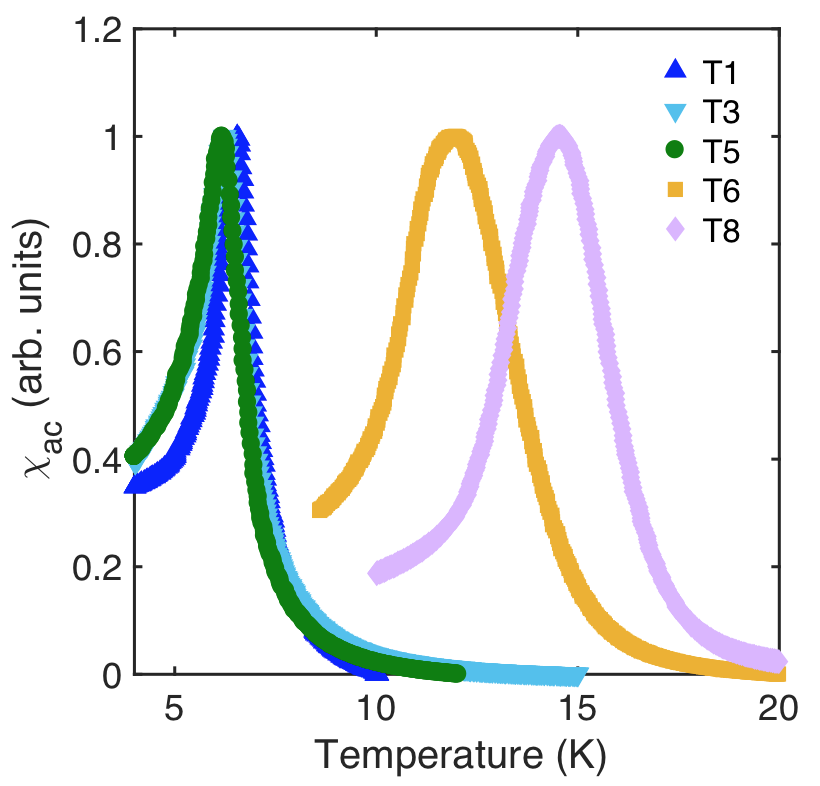}
\caption{Temperature dependence of the AC susceptibility obtained with an excitation frequency of 3.63 kHz for samples cut from a large Y$_{1-x}$La$_{x}$TiO$_{3}$ ($x = 0.22$) single crystal.}
\label{fig:YLa_MvsT}
\end{figure}

\begin{figure}
\includegraphics[width=0.45\textwidth]{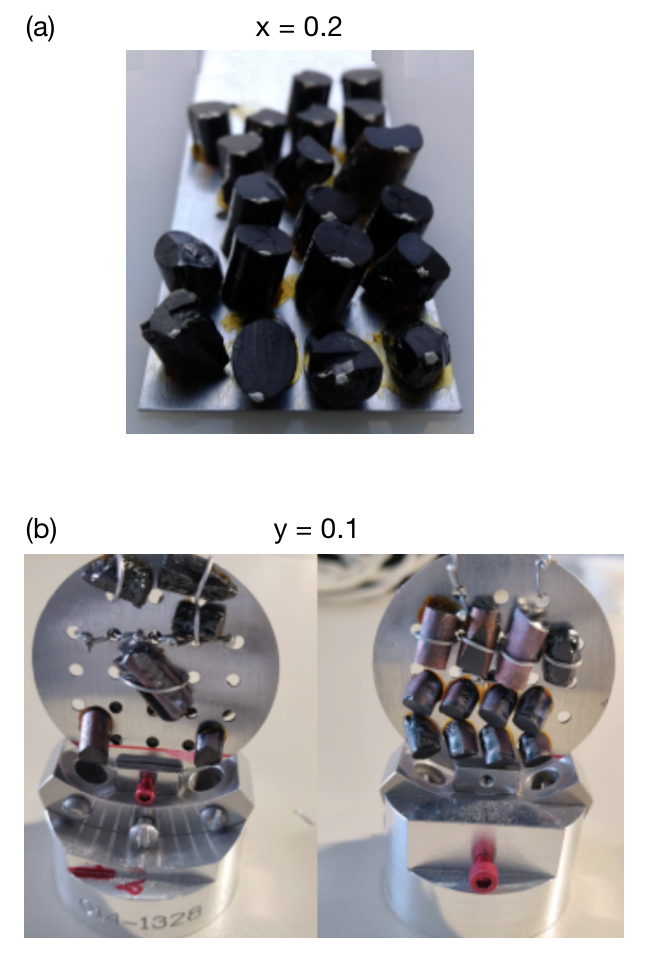}
\caption{Pictures of co-mounted single crystals of (a) Y$_{1-x}$La$_{x}$TiO$_{3}$ ($x = 0.2$) and (b) Y$_{1-y}$Ca$_{y}$TiO$_{3}$ ($y = 0.1$) for inelastic magnetic neutron scattering measurements.}
\label{fig:YLCTO_crystal}
\end{figure}

\begin{figure}
\includegraphics[width=0.4\textwidth]{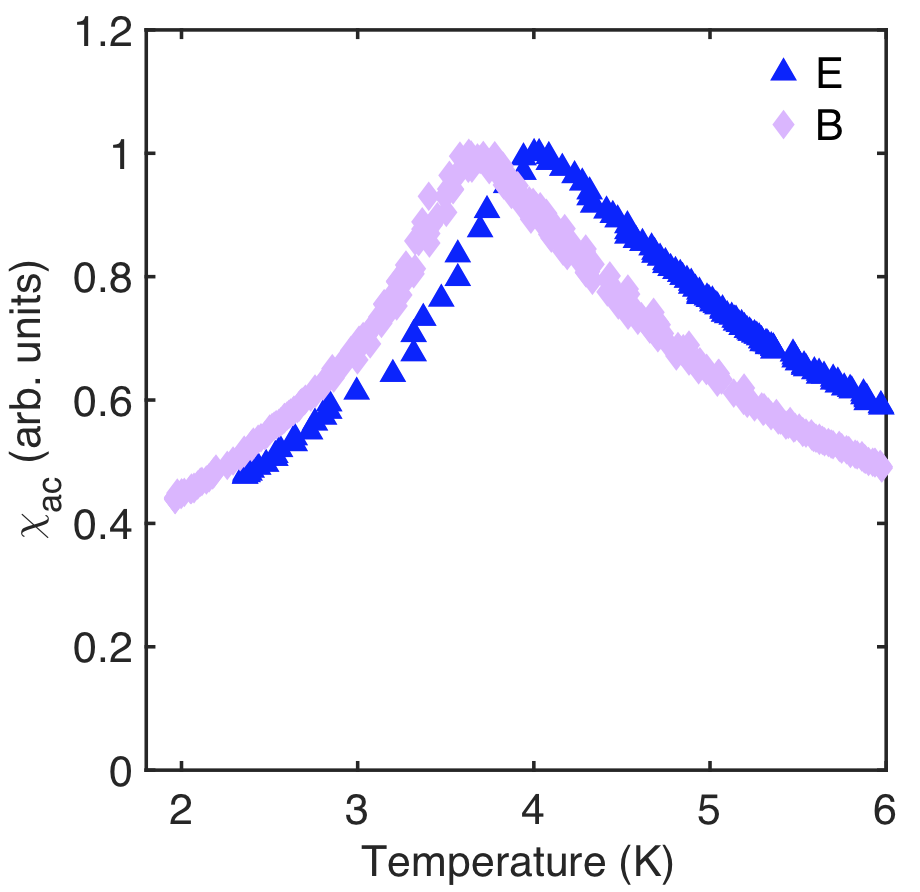}
\caption{Temperature dependence of the AC susceptibility obtained with an excitation frequency of 3.63 kHz for two different samples from the beginning (B) and end (E) of the growth of a large Y$_{1-y}$Ca$_{y}$TiO$_{3}$ ($y = 0.15$) single crystal.}
\label{fig:YCa_MvsT}
\end{figure}

\begin{figure*}
\includegraphics[width=0.95\textwidth]{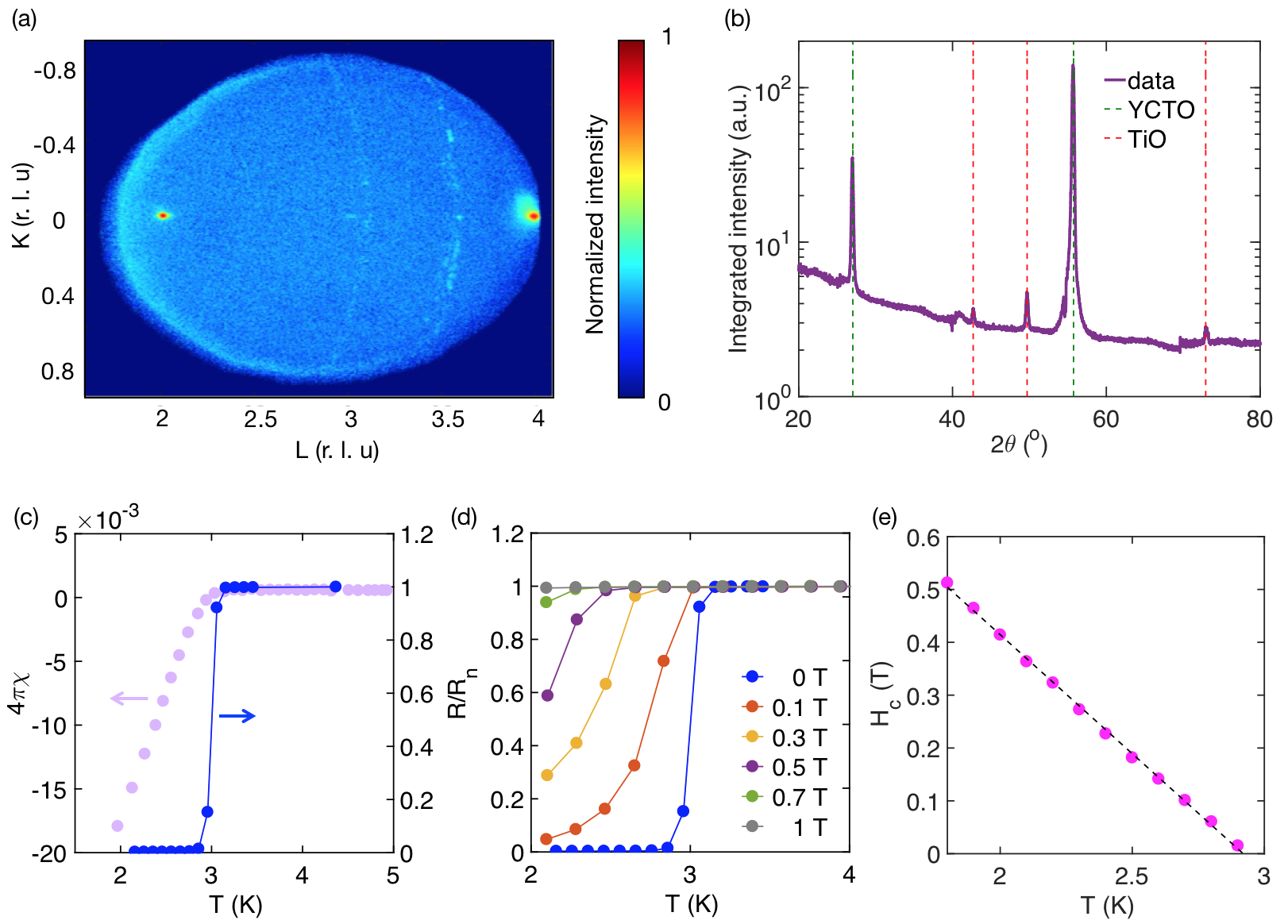}
\caption{(a) Reciprocal space map of the x-ray diffraction intensity in the (0$K$$L$) plane for a Y$_{1-y}$Ca$_{y}$TiO$_{3}$ (YCTO) single crystal with $y = 0.35$. Apart from the strong Bragg peaks at even integer valued $K$ and $L$, powder rings due to a polycrystalline impurity phase are visible. The intensity has been normalized to the peak intensity at (002)$_\text{o}$ (b) Integrated XRD intensity as a function of the scattering angle $2\theta$ for the same $y = 0.35$ crystal as in (a). The calculated $2\theta$ values along with those for the TiO impurity phase are indicated as dashed lines. (c) Temperature dependence of the superconducting volume fraction $4\pi\chi$ obtained from zero-field-cooled magnetization in an applied $c$-axis magnetic field of 50 Oe (left), and of the sample resistance $R$ (normalized to the value in the normal state, $R_n$). The onset temperatures are nearly indistinguishable. The data are for a $y = 0.4$ sample. (d,e) Temperature dependence of the normalized resistance in different applied magnetic fields (d) and of the critical magnetic field (e) for the same sample as in (c).}
\label{fig:xrd}
\end{figure*}

For the growth of Y$_{1-x}$La$_{x}$TiO$_{3}$ single crystals, we fixed the value of $\delta$ at 0.04, $i.e.$, the same value as for YTiO$_3$ in order to avoid property changes due to differences in oxygen off-stoichiometry in the {\it starting} material composition \cite{HameedYLa2021}. Intriguingly, however, we found that, at large $x$, the changes in $T_C$ along the growth axis can be particularly severe. We present here the characterization of a crystal with $x = 0.22$. Similar to the YTiO$_3$ single crystal in Fig.~\ref{fig:YTOSC}, a 4 cm long crystal was sectioned into approximately equal parts (eight instead of four, in this case). As before, T1 refers to the sample at the end of the growth; T8 refers to the sample near the beginning of the growth. 

Figure~\ref{fig:YLa_MvsT} shows the temperature dependence of the AC magnetic susceptibility for five such samples. $T_C$, defined here as the temperature of the susceptibility peak, is summarized in Table~\ref{tab:table2}. Samples T1, T3 and T5 show similar $T_C$ values, whereas samples T6 and T8 exhibit Curie temperatures that are more than a factor of two higher. Similar to the YTiO$_3$ characterization, we also measured the chemical composition of samples across the length of the growth. This is summarized in Table~\ref{tab:table2}. Clearly, no significant differences are observed in the cation ratios, which brings us to the conclusion that the significant differences in magnetic properties are likely associated with oxygen interstitials. This is perhaps not surprising, given the extreme sensitivity of the magnetic and transport properties of LaTiO$_3$ to oxygen interstitials \cite{Meijer1999}. Yet, similar to YTiO$_3$, for sufficiently long growths, we found that an extended region of nearly constant $T_C$ exists towards the latter part of the growth. However, as demonstrated in the present work, careful characterization needs to be performed to minimize the effects of such overoxidation-related inhomogeneities, particularly when selecting large samples for neutron measurements. Note that unlike the case of YTiO$_3$, the `stable' latter part of the growth has a lower $T_C$ compared with the rest. For our own inelastic neutron measurements of the spin-wave spectra at large La-substitution levels, we used samples from the latter parts of the growths that exhibit a nearly constant $T_C$. Given the strong suppression of the ordered moments exhibited by these samples with La substitution \cite{HameedYLa2021}, it was necessary to co-mount samples from many different growths, in order to obtain a sufficiently large signal-to-noise ratio (see Fig.~\ref{fig:YLCTO_crystal}) \cite{HameedSW2021}.

%MG: Here and earlier in the paper, cite the spin-wave paper as manuscript in preparation. Were these large samples also used for order-parameter measurements, or were those done on individual crystals? Perhaps this could be clarified in the neutron characterization section.

\begin{table}[h!]
  \begin{center}
    \caption{Chemical composition and Curie temperatures of the Y$_{1-x}$La$_{x}$TiO$_{3}$ single crystal samples with $x = 0.22$.}
    \label{tab:table2}
    \vspace*{5mm}

    \begin{tabular}{c|c|c|c}
       \textbf{Sample} & \textbf{x} & \textbf{(Y+La):Ti} &  \textbf{$T_C$ (K)}  \\

      \hline
      T1 & 0.236(2) & 0.978(18) & 6.5(1) \\
      T3 & 0.232(2) & 0.983(12) & 6.2(1) \\
      T5 & - & 0.983(14) & 6.2(1) \\
      T6 & 0.232(1) & 0.984(4) & 12.0(5) \\
      T8 & 0.236(4) & 0.976(9) & 14.5(3) \\
    \end{tabular}
  \end{center}
\end{table}

\subsection{Characterization of Y$_{1-y}$Ca$_{y}$TiO$_{3}$ single crystals}
\label{section:YCTO}

The Y$_{1-y}$Ca$_{y}$TiO$_{3}$ growth and characterization results differed from those for Y$_{1-x}$La$_{x}$TiO$_{3}$ in a number of ways. First, several attempts to grow single crystals with $y = 0.05$ and $\delta = 0.04$ failed and yielded only polycrystalline material. In order to obtain single crystals, it was found to be necessary to decrease $\delta$ with increasing $y$. Part of the reason for this is most likely due to the increased melt temperature of the Ca-doped variant, which apparently leads to a stronger oxidizing effect \cite{Roth2008}. The values of $\delta$ used ranged from 0.04 for $y = 0$ to 0.12 for $y = 0.5$. Unlike the La-substituted system, however, Curie-temperature measurements revealed no significant inhomogeneity along large single crystals of Y$_{1-y}$Ca$_{y}$TiO$_{3}$, even at high doping levels. As an example, Fig.~\ref{fig:YCa_MvsT} shows the temperature dependence of the AC susceptibility for two samples from a large ($\sim$ 4 cm long) $y = 0.15$ single crystal. Here, sample `B' is from a region at the beginning of the growth, whereas sample `E' is from a region at the end of the growth. Only a small difference in $T_C$ ($\sim$ 0.3 K, $i.e.$, less than 10\%), identified as the position of the susceptibility peak, is observed.

At high Ca-doping levels $y \geq 0.35$, a polycrystalline impurity phase was seen to appear, as revealed by powder rings in XRD measurements (Fig.~\ref{fig:xrd} (a)). The resultant impurity peaks match those of cubic TiO with a lattice parameter of 4.260(3) \AA (Fig.~\ref{fig:xrd} (b)). This particular impurity phase exhibits a superconducting (SC) transition, as discerned from the diamagnetic signal in magnetometry measurements (Fig.~\ref{fig:xrd} (c)). 
Interestingly, although the low-temperature volume fraction $4\pi\chi$ of the SC phase was estimated to be less than 2\%, in some samples, transport measurements (using a van der Pauw geometry) revealed a complete transition in one of the directions into a zero-resistance state (Fig.~\ref{fig:xrd} (c)). This is likely due to the fact that, in the van der Pauw geometry, the equipotential lines originating from the voltage measurement contacts are skewed, unlike in the case of a stripe geometry, which ensures that small SC regions are captured by the voltage measurement leads \cite{DeVries1995,HameedPelc2020}. This allowed us to characterize the SC properties of the TiO phase (Fig.~\ref{fig:xrd} (c)-(e)). We find a SC transition temperature of about 3 K, as defined by the midpoint of the transition (Fig.~\ref{fig:xrd} (c)). In the literature, the SC transition temperature is known to vary from $\sim 0.5 - 7.4$ K depending on the growth \cite{Wang2017,Zhang2017}. 
%MG: How does this value compare to published bulk/film values? - this varies significantly depending on the Ti:O ratio. Therefore, its hard to make a direct comparison. The highest reported in the bulk is ~ 7 K.
Interestingly, the resistive transition in Fig.~\ref{fig:xrd} (c) is sharper than in recent reports for nominally pure TiO in crystalline and film forms \cite{Wang2017,Zhang2017}. This is likely due to the difference in contact geometry, since the prior work used the standard four-probe technique which averages over the sample \cite{Wang2017,Zhang2017}. We also determined the temperature dependence of the critical magnetic field (Fig.~\ref{fig:xrd} (d,e)). The critical field was taken as the field at which the resistance recovers to 50\% of the normal-state value. We observe a linear temperature dependence down to the lowest measured temperature (1.8 K), in agreement with recent work \cite{Wang2017}.

\begin{figure}
\includegraphics[width=0.4\textwidth]{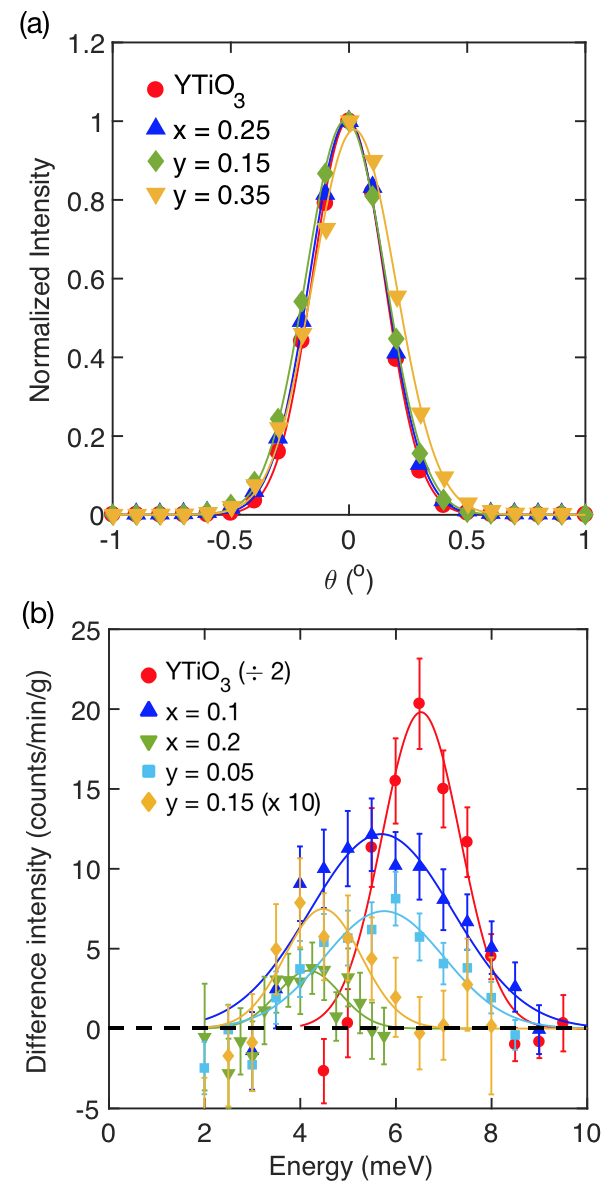}
\caption{(a) Neutron diffraction rocking scans across the (020)$_\text{o}$ Bragg peak for large single crystals of Y$_{1-x}$La$_{x}$TiO$_{3}$ and Y$_{1-y}$Ca$_{y}$TiO$_{3}$. The intensities are normalized to the peak values. The solid lines are gaussian fits to the data. (b) Inelastic neutron scattering measurement of spin-waves at (001)$_\text{o}$. The high-temperature ($T > T_C$) paramagnetic scattering was subtracted from data taken in the ferromagnetically-ordered state at 1.5 K. 
Note that some of the data are scaled, as indicated. The errorbars correspond to one standard deviation. The solid lines are guides to the eye.}
\label{fig:neutron}
\end{figure}

\subsection{Neutron scattering measurements}
\label{section:Neutron}

As noted, a major goal of the present work was to grow sizable single crystals of Y$_{1-x}$La$_{x}$TiO$_{3}$ and Y$_{1-y}$Ca$_{y}$TiO$_{3}$ for magnetic neutron scattering studies. Here we present neutron scattering data that demonstrate the good quality of our crystals. Figure~\ref{fig:neutron}(a) displays rocking scans across the (020)$_\text{o}$ nuclear Bragg peak for single crystals with different chemical compositions. The volumes of the single crystals used were in the range 0.7 - 0.8 cm$^3$ with a mass of $\sim$ 4 g. Clearly, the crystal mosaic, defined here as the full-width-at-half-maximum (FWHM) of gaussian fits, is $<$ 0.5$^\text{o}$, indicative of high crystalline quality. Note that part of the broadening is associated with the instrumental resolution. Therefore, the FWHM represents an upper bound on the crystal mosaic.
%MG: What were the crystal sizes/masses here? Presumably, you are showing data for individual crystals and not for co-mounted samples.
Figure~\ref{fig:neutron}(b) displays spin-wave data for several single crystals. We observe a strong suppression of the spin-wave intensity, both with increasing La substitution and increasing Ca doping. In order to obtain a good signal-noise-ratio for $x \geq 0.2$ and $y \geq 0.1$, we co-mounted $\sim 20$ crystals with total masses in the range 12 - 15 g, as shown in Fig.~\ref{fig:YLCTO_crystal}.
Detailed spin-wave measurements will be reported elsewhere \cite{HameedSW2021}.
%MG: This section is rather short. Please think about adding more text while we wait for feedback from our collaborators.

%-------------------------------------------------------------------
%Conclusions
%-------------------------------------------------------------------

\section{Conclusions}
\label{section:conc}

To summarize, we have successfully grown large single crystals of YTiO$_3$ and its La-substituted and Ca-doped variants. We found that inhomogeneous overoxidation can appear along large single crystals and strongly affects the magnetic and transport properties. These findings indicate that great care must be taken in preparing samples for experimental studies aimed to understand the phase diagram. We have demonstrated that it is possible to obtain elastic and inelastic magnetic neutron scattering data with good signal-to-noise ratios. We have also found a superconducting TiO polycrystalline impurity phase at large Ca-substitution levels. The superconducting transition temperature of this phase is about 3 K, and the critical magnetic field exhibits a linear temperature dependence, consistent with prior results for bulk and film samples. 

%MG: Add proper ORNL and NIST acknowledgements
\section{Acknowledgements}
We thank K. Olson for help with the growth of the Y$_{1-x}$La$_{x}$TiO$_{3}$ ($x=0.22$) crystal and A. Najev for helpful comments on the manuscript. The work at University of Minnesota was funded by the Department of Energy through the University of Minnesota Center for Quantum Materials, under Grant No. DE-SC0016371. Parts of this work were carried out in the University of Minnesota Characterization Facility, which receives partial support from NSF through the MRSEC program. Electron microprobe analysis of the crystal chemical composition was carried out at the Electron Microprobe Laboratory, Department of Earth Sciences, University of Minnesota-Twin Cities. This research used resources of the Advanced Photon Source, a U.S. Department of Energy (DOE) Office of Science User Facility operated for the DOE Office of Science by Argonne National Laboratory under Contract No. DE-AC02-06CH11357. A portion of this research used resources at the High Flux Isotope Reactor, a DOE Office of Science User Facility operated by the Oak Ridge National Laboratory. We acknowledge the support of the National Institute of Standards and Technology, U.S. Department of Commerce, in providing the neutron research facilities used in this work. Certain commercial equipment, instruments, or materials are identified in this paper in order to specify the experimental procedure adequately. Such identification is not intended to imply recommendation or endorsement by the National Institute of Standards and Technology, nor is it intended to imply that the materials or equipment identified are necessarily the best available for the purpose.

%\nocite{*}
%\bibliographystyle{apsrev}
\bibliography{YTOGrowth.bib}

%merlin.mbs apsrev4-1.bst 2010-07-25 4.21a (PWD, AO, DPC) hacked
%Control: key (0)
%Control: author (72) initials jnrlst
%Control: editor formatted (1) identically to author
%Control: production of article title (-1) disabled
%Control: page (0) single
%Control: year (1) truncated
%Control: production of eprint (0) enabled
\begin{thebibliography}{31}%
\makeatletter
\providecommand \@ifxundefined [1]{%
 \@ifx{#1\undefined}
}%
\providecommand \@ifnum [1]{%
 \ifnum #1\expandafter \@firstoftwo
 \else \expandafter \@secondoftwo
 \fi
}%
\providecommand \@ifx [1]{%
 \ifx #1\expandafter \@firstoftwo
 \else \expandafter \@secondoftwo
 \fi
}%
\providecommand \natexlab [1]{#1}%
\providecommand \enquote  [1]{``#1''}%
\providecommand \bibnamefont  [1]{#1}%
\providecommand \bibfnamefont [1]{#1}%
\providecommand \citenamefont [1]{#1}%
\providecommand \href@noop [0]{\@secondoftwo}%
\providecommand \href [0]{\begingroup \@sanitize@url \@href}%
\providecommand \@href[1]{\@@startlink{#1}\@@href}%
\providecommand \@@href[1]{\endgroup#1\@@endlink}%
\providecommand \@sanitize@url [0]{\catcode `\\12\catcode `\$12\catcode
  `\&12\catcode `\#12\catcode `\^12\catcode `\_12\catcode `\%12\relax}%
\providecommand \@@startlink[1]{}%
\providecommand \@@endlink[0]{}%
\providecommand \url  [0]{\begingroup\@sanitize@url \@url }%
\providecommand \@url [1]{\endgroup\@href {#1}{\urlprefix }}%
\providecommand \urlprefix  [0]{URL }%
\providecommand \Eprint [0]{\href }%
\providecommand \doibase [0]{http://dx.doi.org/}%
\providecommand \selectlanguage [0]{\@gobble}%
\providecommand \bibinfo  [0]{\@secondoftwo}%
\providecommand \bibfield  [0]{\@secondoftwo}%
\providecommand \translation [1]{[#1]}%
\providecommand \BibitemOpen [0]{}%
\providecommand \bibitemStop [0]{}%
\providecommand \bibitemNoStop [0]{.\EOS\space}%
\providecommand \EOS [0]{\spacefactor3000\relax}%
\providecommand \BibitemShut  [1]{\csname bibitem#1\endcsname}%
\let\auto@bib@innerbib\@empty
%</preamble>
\bibitem [{\citenamefont {Mochizuki}\ and\ \citenamefont
  {Imada}(2004)}]{Mochizuki2004}%
  \BibitemOpen
  \bibfield  {author} {\bibinfo {author} {\bibfnamefont {M.}~\bibnamefont
  {Mochizuki}}\ and\ \bibinfo {author} {\bibfnamefont {M.}~\bibnamefont
  {Imada}},\ }\href@noop {} {\bibfield  {journal} {\bibinfo  {journal} {New J.
  Phys.}\ }\textbf {\bibinfo {volume} {6}},\ \bibinfo {pages} {154} (\bibinfo
  {year} {2004})}\BibitemShut {NoStop}%
\bibitem [{\citenamefont {MacLean}\ \emph {et~al.}(1979)\citenamefont
  {MacLean}, \citenamefont {Ng},\ and\ \citenamefont {Greedan}}]{MacLean1979}%
  \BibitemOpen
  \bibfield  {author} {\bibinfo {author} {\bibfnamefont {D.~A.}\ \bibnamefont
  {MacLean}}, \bibinfo {author} {\bibfnamefont {H.~N.}\ \bibnamefont {Ng}}, \
  and\ \bibinfo {author} {\bibfnamefont {J.~E.}\ \bibnamefont {Greedan}},\
  }\href@noop {} {\bibfield  {journal} {\bibinfo  {journal} {J. Solid State
  Chem.}\ }\textbf {\bibinfo {volume} {30}},\ \bibinfo {pages} {35} (\bibinfo
  {year} {1979})}\BibitemShut {NoStop}%
\bibitem [{\citenamefont {Greedan}(1985)}]{Greedan1985}%
  \BibitemOpen
  \bibfield  {author} {\bibinfo {author} {\bibfnamefont {J.~E.}\ \bibnamefont
  {Greedan}},\ }\href@noop {} {\bibfield  {journal} {\bibinfo  {journal} {J.
  Less Common Metals}\ }\textbf {\bibinfo {volume} {111}},\ \bibinfo {pages}
  {335} (\bibinfo {year} {1985})}\BibitemShut {NoStop}%
\bibitem [{\citenamefont {Zhou}\ and\ \citenamefont
  {Goodenough}(2005{\natexlab{a}})}]{Zhou2005a}%
  \BibitemOpen
  \bibfield  {author} {\bibinfo {author} {\bibfnamefont {H.~D.}\ \bibnamefont
  {Zhou}}\ and\ \bibinfo {author} {\bibfnamefont {J.~B.}\ \bibnamefont
  {Goodenough}},\ }\href@noop {} {\bibfield  {journal} {\bibinfo  {journal} {J.
  Phys. Condens. Matter}\ }\textbf {\bibinfo {volume} {17}},\ \bibinfo {pages}
  {7395} (\bibinfo {year} {2005}{\natexlab{a}})}\BibitemShut {NoStop}%
\bibitem [{\citenamefont {Goral}\ \emph {et~al.}(1982)\citenamefont {Goral},
  \citenamefont {Greedan},\ and\ \citenamefont {MacLean}}]{Goral1982}%
  \BibitemOpen
  \bibfield  {author} {\bibinfo {author} {\bibfnamefont {J.~P.}\ \bibnamefont
  {Goral}}, \bibinfo {author} {\bibfnamefont {J.~E.}\ \bibnamefont {Greedan}},
  \ and\ \bibinfo {author} {\bibfnamefont {D.~A.}\ \bibnamefont {MacLean}},\
  }\href@noop {} {\bibfield  {journal} {\bibinfo  {journal} {J. Solid State
  Chem.}\ }\textbf {\bibinfo {volume} {43}},\ \bibinfo {pages} {244} (\bibinfo
  {year} {1982})}\BibitemShut {NoStop}%
\bibitem [{\citenamefont {Okimoto}\ \emph {et~al.}(1995)\citenamefont
  {Okimoto}, \citenamefont {Katsufuji}, \citenamefont {Okada}, \citenamefont
  {Arima},\ and\ \citenamefont {Tokura}}]{Okimoto1995}%
  \BibitemOpen
  \bibfield  {author} {\bibinfo {author} {\bibfnamefont {Y.}~\bibnamefont
  {Okimoto}}, \bibinfo {author} {\bibfnamefont {T.}~\bibnamefont {Katsufuji}},
  \bibinfo {author} {\bibfnamefont {Y.}~\bibnamefont {Okada}}, \bibinfo
  {author} {\bibfnamefont {T.}~\bibnamefont {Arima}}, \ and\ \bibinfo {author}
  {\bibfnamefont {Y.}~\bibnamefont {Tokura}},\ }\href@noop {} {\bibfield
  {journal} {\bibinfo  {journal} {Phys. Rev. B}\ }\textbf {\bibinfo {volume}
  {51}},\ \bibinfo {pages} {9581} (\bibinfo {year} {1995})}\BibitemShut
  {NoStop}%
\bibitem [{\citenamefont {Zhou}\ and\ \citenamefont
  {Goodenough}(2005{\natexlab{b}})}]{Zhou2005}%
  \BibitemOpen
  \bibfield  {author} {\bibinfo {author} {\bibfnamefont {H.~D.}\ \bibnamefont
  {Zhou}}\ and\ \bibinfo {author} {\bibfnamefont {J.~B.}\ \bibnamefont
  {Goodenough}},\ }\href@noop {} {\bibfield  {journal} {\bibinfo  {journal}
  {Phys. Rev. B}\ }\textbf {\bibinfo {volume} {71}},\ \bibinfo {pages} {184431}
  (\bibinfo {year} {2005}{\natexlab{b}})}\BibitemShut {NoStop}%
\bibitem [{\citenamefont {Hameed}\ \emph
  {et~al.}(2021{\natexlab{a}})\citenamefont {Hameed}, \citenamefont
  {El-Khatib}, \citenamefont {Olson}, \citenamefont {Yu}, \citenamefont
  {Williams}, \citenamefont {Hong}, \citenamefont {Sheng}, \citenamefont
  {Yamakawa}, \citenamefont {Zang}, \citenamefont {Uemura}, \citenamefont
  {Zhao}, \citenamefont {Jin}, \citenamefont {Fu}, \citenamefont {Gu},
  \citenamefont {Ning}, \citenamefont {Cai}, \citenamefont {Kojima},
  \citenamefont {Freeland}, \citenamefont {Matsuda}, \citenamefont {Leighton},\
  and\ \citenamefont {Greven}}]{HameedYLa2021}%
  \BibitemOpen
  \bibfield  {author} {\bibinfo {author} {\bibfnamefont {S.}~\bibnamefont
  {Hameed}}, \bibinfo {author} {\bibfnamefont {S.}~\bibnamefont {El-Khatib}},
  \bibinfo {author} {\bibfnamefont {K.~P.}\ \bibnamefont {Olson}}, \bibinfo
  {author} {\bibfnamefont {B.}~\bibnamefont {Yu}}, \bibinfo {author}
  {\bibfnamefont {T.~J.}\ \bibnamefont {Williams}}, \bibinfo {author}
  {\bibfnamefont {T.}~\bibnamefont {Hong}}, \bibinfo {author} {\bibfnamefont
  {Q.}~\bibnamefont {Sheng}}, \bibinfo {author} {\bibfnamefont
  {K.}~\bibnamefont {Yamakawa}}, \bibinfo {author} {\bibfnamefont
  {J.}~\bibnamefont {Zang}}, \bibinfo {author} {\bibfnamefont {Y.~J.}\
  \bibnamefont {Uemura}}, \bibinfo {author} {\bibfnamefont {G.~Q.}\
  \bibnamefont {Zhao}}, \bibinfo {author} {\bibfnamefont {C.~Q.}\ \bibnamefont
  {Jin}}, \bibinfo {author} {\bibfnamefont {L.}~\bibnamefont {Fu}}, \bibinfo
  {author} {\bibfnamefont {Y.}~\bibnamefont {Gu}}, \bibinfo {author}
  {\bibfnamefont {F.}~\bibnamefont {Ning}}, \bibinfo {author} {\bibfnamefont
  {Y.}~\bibnamefont {Cai}}, \bibinfo {author} {\bibfnamefont {K.~M.}\
  \bibnamefont {Kojima}}, \bibinfo {author} {\bibfnamefont {J.~W.}\
  \bibnamefont {Freeland}}, \bibinfo {author} {\bibfnamefont {M.}~\bibnamefont
  {Matsuda}}, \bibinfo {author} {\bibfnamefont {C.}~\bibnamefont {Leighton}}, \
  and\ \bibinfo {author} {\bibfnamefont {M.}~\bibnamefont {Greven}},\
  }\href@noop {} {\bibfield  {journal} {\bibinfo  {journal} {arXiv:2103.08565}\
  } (\bibinfo {year} {2021}{\natexlab{a}})}\BibitemShut {NoStop}%
\bibitem [{\citenamefont {Amow}\ \emph {et~al.}(2000)\citenamefont {Amow},
  \citenamefont {Zhou},\ and\ \citenamefont {Goodenough}}]{Amow2000}%
  \BibitemOpen
  \bibfield  {author} {\bibinfo {author} {\bibfnamefont {G.}~\bibnamefont
  {Amow}}, \bibinfo {author} {\bibfnamefont {J.-S.}\ \bibnamefont {Zhou}}, \
  and\ \bibinfo {author} {\bibfnamefont {J.~B.}\ \bibnamefont {Goodenough}},\
  }\href@noop {} {\bibfield  {journal} {\bibinfo  {journal} {J. Solid State
  Chem.}\ }\textbf {\bibinfo {volume} {154}},\ \bibinfo {pages} {619} (\bibinfo
  {year} {2000})}\BibitemShut {NoStop}%
\bibitem [{\citenamefont {Tokura}\ \emph
  {et~al.}(1993{\natexlab{a}})\citenamefont {Tokura}, \citenamefont {Taguchi},
  \citenamefont {Okada}, \citenamefont {Fujishima}, \citenamefont {Arima},
  \citenamefont {Kumagai},\ and\ \citenamefont {Iye}}]{Tokura1993La}%
  \BibitemOpen
  \bibfield  {author} {\bibinfo {author} {\bibfnamefont {Y.}~\bibnamefont
  {Tokura}}, \bibinfo {author} {\bibfnamefont {Y.}~\bibnamefont {Taguchi}},
  \bibinfo {author} {\bibfnamefont {Y.}~\bibnamefont {Okada}}, \bibinfo
  {author} {\bibfnamefont {Y.}~\bibnamefont {Fujishima}}, \bibinfo {author}
  {\bibfnamefont {T.}~\bibnamefont {Arima}}, \bibinfo {author} {\bibfnamefont
  {K.}~\bibnamefont {Kumagai}}, \ and\ \bibinfo {author} {\bibfnamefont
  {Y.}~\bibnamefont {Iye}},\ }\href@noop {} {\bibfield  {journal} {\bibinfo
  {journal} {Phys. Rev. Lett.}\ }\textbf {\bibinfo {volume} {70}},\ \bibinfo
  {pages} {2126} (\bibinfo {year} {1993}{\natexlab{a}})}\BibitemShut {NoStop}%
\bibitem [{\citenamefont {Tokura}\ \emph
  {et~al.}(1993{\natexlab{b}})\citenamefont {Tokura}, \citenamefont {Taguchi},
  \citenamefont {Moritomo}, \citenamefont {Kumagai}, \citenamefont {Suzuki},\
  and\ \citenamefont {Iye}}]{Tokura1993}%
  \BibitemOpen
  \bibfield  {author} {\bibinfo {author} {\bibfnamefont {Y.}~\bibnamefont
  {Tokura}}, \bibinfo {author} {\bibfnamefont {Y.}~\bibnamefont {Taguchi}},
  \bibinfo {author} {\bibfnamefont {Y.}~\bibnamefont {Moritomo}}, \bibinfo
  {author} {\bibfnamefont {K.}~\bibnamefont {Kumagai}}, \bibinfo {author}
  {\bibfnamefont {T.}~\bibnamefont {Suzuki}}, \ and\ \bibinfo {author}
  {\bibfnamefont {Y.}~\bibnamefont {Iye}},\ }\href@noop {} {\bibfield
  {journal} {\bibinfo  {journal} {Phys. Rev. B}\ }\textbf {\bibinfo {volume}
  {48}},\ \bibinfo {pages} {14063} (\bibinfo {year}
  {1993}{\natexlab{b}})}\BibitemShut {NoStop}%
\bibitem [{\citenamefont {Hameed}\ \emph
  {et~al.}(2021{\natexlab{b}})\citenamefont {Hameed}, \citenamefont {Joe},
  \citenamefont {Gautreau}, \citenamefont {Freeland}, \citenamefont {Birol},\
  and\ \citenamefont {Greven}}]{HameedYCa2021}%
  \BibitemOpen
  \bibfield  {author} {\bibinfo {author} {\bibfnamefont {S.}~\bibnamefont
  {Hameed}}, \bibinfo {author} {\bibfnamefont {J.}~\bibnamefont {Joe}},
  \bibinfo {author} {\bibfnamefont {D.~M.}\ \bibnamefont {Gautreau}}, \bibinfo
  {author} {\bibfnamefont {J.~W.}\ \bibnamefont {Freeland}}, \bibinfo {author}
  {\bibfnamefont {T.}~\bibnamefont {Birol}}, \ and\ \bibinfo {author}
  {\bibfnamefont {M.}~\bibnamefont {Greven}},\ }\href@noop {} {\bibfield
  {journal} {\bibinfo  {journal} {arXiv:2103.08566}\ } (\bibinfo {year}
  {2021}{\natexlab{b}})}\BibitemShut {NoStop}%
\bibitem [{\citenamefont {Xu}\ \emph {et~al.}(2016{\natexlab{a}})\citenamefont
  {Xu}, \citenamefont {Droubay}, \citenamefont {Jeong}, \citenamefont
  {Mkhoyan}, \citenamefont {Sushko}, \citenamefont {Chambers},\ and\
  \citenamefont {Jalan}}]{Xu2016a}%
  \BibitemOpen
  \bibfield  {author} {\bibinfo {author} {\bibfnamefont {P.}~\bibnamefont
  {Xu}}, \bibinfo {author} {\bibfnamefont {T.~C.}\ \bibnamefont {Droubay}},
  \bibinfo {author} {\bibfnamefont {J.~S.}\ \bibnamefont {Jeong}}, \bibinfo
  {author} {\bibfnamefont {K.~A.}\ \bibnamefont {Mkhoyan}}, \bibinfo {author}
  {\bibfnamefont {P.~V.}\ \bibnamefont {Sushko}}, \bibinfo {author}
  {\bibfnamefont {S.~A.}\ \bibnamefont {Chambers}}, \ and\ \bibinfo {author}
  {\bibfnamefont {B.}~\bibnamefont {Jalan}},\ }\href@noop {} {\bibfield
  {journal} {\bibinfo  {journal} {Adv. Mater. Interfaces}\ }\textbf {\bibinfo
  {volume} {3}},\ \bibinfo {pages} {1500432} (\bibinfo {year}
  {2016}{\natexlab{a}})}\BibitemShut {NoStop}%
\bibitem [{\citenamefont {Xu}\ \emph {et~al.}(2016{\natexlab{b}})\citenamefont
  {Xu}, \citenamefont {Ayino}, \citenamefont {Cheng}, \citenamefont {Pribiag},
  \citenamefont {Comes}, \citenamefont {Sushko}, \citenamefont {Chambers},\
  and\ \citenamefont {Jalan}}]{Xu2016b}%
  \BibitemOpen
  \bibfield  {author} {\bibinfo {author} {\bibfnamefont {P.}~\bibnamefont
  {Xu}}, \bibinfo {author} {\bibfnamefont {Y.}~\bibnamefont {Ayino}}, \bibinfo
  {author} {\bibfnamefont {C.}~\bibnamefont {Cheng}}, \bibinfo {author}
  {\bibfnamefont {V.~S.}\ \bibnamefont {Pribiag}}, \bibinfo {author}
  {\bibfnamefont {R.~B.}\ \bibnamefont {Comes}}, \bibinfo {author}
  {\bibfnamefont {P.~V.}\ \bibnamefont {Sushko}}, \bibinfo {author}
  {\bibfnamefont {S.~A.}\ \bibnamefont {Chambers}}, \ and\ \bibinfo {author}
  {\bibfnamefont {B.}~\bibnamefont {Jalan}},\ }\href@noop {} {\bibfield
  {journal} {\bibinfo  {journal} {Phys. Rev. Lett.}\ }\textbf {\bibinfo
  {volume} {117}},\ \bibinfo {pages} {106803} (\bibinfo {year}
  {2016}{\natexlab{b}})}\BibitemShut {NoStop}%
\bibitem [{\citenamefont {Cao}\ \emph {et~al.}(2016)\citenamefont {Cao},
  \citenamefont {Yang}, \citenamefont {Kareev}, \citenamefont {Liu},
  \citenamefont {Meyers}, \citenamefont {Middey}, \citenamefont {Choudhury},
  \citenamefont {Shafer}, \citenamefont {Guo}, \citenamefont {Freeland},
  \citenamefont {Arenholz}, \citenamefont {Gu},\ and\ \citenamefont
  {Chakhalian}}]{Cao2016a}%
  \BibitemOpen
  \bibfield  {author} {\bibinfo {author} {\bibfnamefont {Y.}~\bibnamefont
  {Cao}}, \bibinfo {author} {\bibfnamefont {Z.}~\bibnamefont {Yang}}, \bibinfo
  {author} {\bibfnamefont {M.}~\bibnamefont {Kareev}}, \bibinfo {author}
  {\bibfnamefont {X.}~\bibnamefont {Liu}}, \bibinfo {author} {\bibfnamefont
  {D.}~\bibnamefont {Meyers}}, \bibinfo {author} {\bibfnamefont
  {S.}~\bibnamefont {Middey}}, \bibinfo {author} {\bibfnamefont
  {D.}~\bibnamefont {Choudhury}}, \bibinfo {author} {\bibfnamefont
  {P.}~\bibnamefont {Shafer}}, \bibinfo {author} {\bibfnamefont
  {J.}~\bibnamefont {Guo}}, \bibinfo {author} {\bibfnamefont {J.~W.}\
  \bibnamefont {Freeland}}, \bibinfo {author} {\bibfnamefont {E.}~\bibnamefont
  {Arenholz}}, \bibinfo {author} {\bibfnamefont {L.}~\bibnamefont {Gu}}, \ and\
  \bibinfo {author} {\bibfnamefont {J.}~\bibnamefont {Chakhalian}},\
  }\href@noop {} {\bibfield  {journal} {\bibinfo  {journal} {Phys. Rev. Lett.}\
  }\textbf {\bibinfo {volume} {116}},\ \bibinfo {pages} {076802} (\bibinfo
  {year} {2016})}\BibitemShut {NoStop}%
\bibitem [{\citenamefont {Raghavan}\ \emph {et~al.}(2015)\citenamefont
  {Raghavan}, \citenamefont {Zhang},\ and\ \citenamefont
  {Stemmer}}]{Raghavan2015}%
  \BibitemOpen
  \bibfield  {author} {\bibinfo {author} {\bibfnamefont {S.}~\bibnamefont
  {Raghavan}}, \bibinfo {author} {\bibfnamefont {J.~Y.}\ \bibnamefont {Zhang}},
  \ and\ \bibinfo {author} {\bibfnamefont {S.}~\bibnamefont {Stemmer}},\
  }\href@noop {} {\bibfield  {journal} {\bibinfo  {journal} {Appl. Phys.
  Lett.}\ }\textbf {\bibinfo {volume} {106}},\ \bibinfo {pages} {132104}
  (\bibinfo {year} {2015})}\BibitemShut {NoStop}%
\bibitem [{\citenamefont {Moetakef}\ \emph {et~al.}(2012)\citenamefont
  {Moetakef}, \citenamefont {Williams}, \citenamefont {Ouellette},
  \citenamefont {Kajdos}, \citenamefont {Goldhaber-Gordon}, \citenamefont
  {Allen},\ and\ \citenamefont {Stemmer}}]{Moetakef2012}%
  \BibitemOpen
  \bibfield  {author} {\bibinfo {author} {\bibfnamefont {P.}~\bibnamefont
  {Moetakef}}, \bibinfo {author} {\bibfnamefont {J.~R.}\ \bibnamefont
  {Williams}}, \bibinfo {author} {\bibfnamefont {D.~G.}\ \bibnamefont
  {Ouellette}}, \bibinfo {author} {\bibfnamefont {A.~P.}\ \bibnamefont
  {Kajdos}}, \bibinfo {author} {\bibfnamefont {D.}~\bibnamefont
  {Goldhaber-Gordon}}, \bibinfo {author} {\bibfnamefont {S.~J.}\ \bibnamefont
  {Allen}}, \ and\ \bibinfo {author} {\bibfnamefont {S.}~\bibnamefont
  {Stemmer}},\ }\href@noop {} {\bibfield  {journal} {\bibinfo  {journal} {Phys.
  Rev. X}\ }\textbf {\bibinfo {volume} {2}},\ \bibinfo {pages} {021014}
  (\bibinfo {year} {2012})}\BibitemShut {NoStop}%
\bibitem [{\citenamefont {Grisolia}\ \emph {et~al.}(2016)\citenamefont
  {Grisolia}, \citenamefont {Varignon}, \citenamefont {Sanchez-Santolino},
  \citenamefont {Arora}, \citenamefont {Valencia}, \citenamefont {Varela},
  \citenamefont {Abrudan}, \citenamefont {Weschke}, \citenamefont {Schierle},
  \citenamefont {Rault}, \citenamefont {Rueff}, \citenamefont
  {Barth{\'{e}}l{\'{e}}my}, \citenamefont {Santamaria},\ and\ \citenamefont
  {Bibes}}]{Grisolia2016}%
  \BibitemOpen
  \bibfield  {author} {\bibinfo {author} {\bibfnamefont {M.~N.}\ \bibnamefont
  {Grisolia}}, \bibinfo {author} {\bibfnamefont {J.}~\bibnamefont {Varignon}},
  \bibinfo {author} {\bibfnamefont {G.}~\bibnamefont {Sanchez-Santolino}},
  \bibinfo {author} {\bibfnamefont {A.}~\bibnamefont {Arora}}, \bibinfo
  {author} {\bibfnamefont {S.}~\bibnamefont {Valencia}}, \bibinfo {author}
  {\bibfnamefont {M.}~\bibnamefont {Varela}}, \bibinfo {author} {\bibfnamefont
  {R.}~\bibnamefont {Abrudan}}, \bibinfo {author} {\bibfnamefont
  {E.}~\bibnamefont {Weschke}}, \bibinfo {author} {\bibfnamefont
  {E.}~\bibnamefont {Schierle}}, \bibinfo {author} {\bibfnamefont {J.~E.}\
  \bibnamefont {Rault}}, \bibinfo {author} {\bibfnamefont {J.~P.}\ \bibnamefont
  {Rueff}}, \bibinfo {author} {\bibfnamefont {A.}~\bibnamefont
  {Barth{\'{e}}l{\'{e}}my}}, \bibinfo {author} {\bibfnamefont {J.}~\bibnamefont
  {Santamaria}}, \ and\ \bibinfo {author} {\bibfnamefont {M.}~\bibnamefont
  {Bibes}},\ }\href@noop {} {\bibfield  {journal} {\bibinfo  {journal} {Nat.
  Phys.}\ }\textbf {\bibinfo {volume} {12}},\ \bibinfo {pages} {484} (\bibinfo
  {year} {2016})}\BibitemShut {NoStop}%
\bibitem [{\citenamefont {Aeschlimann}\ \emph {et~al.}(2018)\citenamefont
  {Aeschlimann}, \citenamefont {Preziosi}, \citenamefont {Scheiderer},
  \citenamefont {Sing}, \citenamefont {Valencia}, \citenamefont {Santamaria},
  \citenamefont {Luo}, \citenamefont {Ryll}, \citenamefont {Radu},
  \citenamefont {Claessen}, \citenamefont {Piamonteze},\ and\ \citenamefont
  {Bibes}}]{Aeschlimann2018}%
  \BibitemOpen
  \bibfield  {author} {\bibinfo {author} {\bibfnamefont {R.}~\bibnamefont
  {Aeschlimann}}, \bibinfo {author} {\bibfnamefont {D.}~\bibnamefont
  {Preziosi}}, \bibinfo {author} {\bibfnamefont {P.}~\bibnamefont
  {Scheiderer}}, \bibinfo {author} {\bibfnamefont {M.}~\bibnamefont {Sing}},
  \bibinfo {author} {\bibfnamefont {S.}~\bibnamefont {Valencia}}, \bibinfo
  {author} {\bibfnamefont {J.}~\bibnamefont {Santamaria}}, \bibinfo {author}
  {\bibfnamefont {C.}~\bibnamefont {Luo}}, \bibinfo {author} {\bibfnamefont
  {H.}~\bibnamefont {Ryll}}, \bibinfo {author} {\bibfnamefont {F.}~\bibnamefont
  {Radu}}, \bibinfo {author} {\bibfnamefont {R.}~\bibnamefont {Claessen}},
  \bibinfo {author} {\bibfnamefont {C.}~\bibnamefont {Piamonteze}}, \ and\
  \bibinfo {author} {\bibfnamefont {M.}~\bibnamefont {Bibes}},\ }\href@noop {}
  {\bibfield  {journal} {\bibinfo  {journal} {Adv. Mater.}\ }\textbf {\bibinfo
  {volume} {30}},\ \bibinfo {pages} {1707489} (\bibinfo {year}
  {2018})}\BibitemShut {NoStop}%
\bibitem [{\citenamefont {Aeschlimann}\ \emph {et~al.}(2021)\citenamefont
  {Aeschlimann}, \citenamefont {Grisolia}, \citenamefont {Sanchez-Santolino},
  \citenamefont {Varignon}, \citenamefont {Choueikani}, \citenamefont
  {Mattana}, \citenamefont {Garcia}, \citenamefont {Fusil}, \citenamefont
  {Fr{\"{o}}hlich}, \citenamefont {Braden}, \citenamefont {Delley},
  \citenamefont {Varela}, \citenamefont {Ohresser}, \citenamefont {Santamaria},
  \citenamefont {Barth{\'{e}}l{\'{e}}my}, \citenamefont {Piamonteze},\ and\
  \citenamefont {Bibes}}]{Aeschlimann2021}%
  \BibitemOpen
  \bibfield  {author} {\bibinfo {author} {\bibfnamefont {R.}~\bibnamefont
  {Aeschlimann}}, \bibinfo {author} {\bibfnamefont {M.~N.}\ \bibnamefont
  {Grisolia}}, \bibinfo {author} {\bibfnamefont {G.}~\bibnamefont
  {Sanchez-Santolino}}, \bibinfo {author} {\bibfnamefont {J.}~\bibnamefont
  {Varignon}}, \bibinfo {author} {\bibfnamefont {F.}~\bibnamefont
  {Choueikani}}, \bibinfo {author} {\bibfnamefont {R.}~\bibnamefont {Mattana}},
  \bibinfo {author} {\bibfnamefont {V.}~\bibnamefont {Garcia}}, \bibinfo
  {author} {\bibfnamefont {S.}~\bibnamefont {Fusil}}, \bibinfo {author}
  {\bibfnamefont {T.}~\bibnamefont {Fr{\"{o}}hlich}}, \bibinfo {author}
  {\bibfnamefont {M.}~\bibnamefont {Braden}}, \bibinfo {author} {\bibfnamefont
  {B.}~\bibnamefont {Delley}}, \bibinfo {author} {\bibfnamefont
  {M.}~\bibnamefont {Varela}}, \bibinfo {author} {\bibfnamefont
  {P.}~\bibnamefont {Ohresser}}, \bibinfo {author} {\bibfnamefont
  {J.}~\bibnamefont {Santamaria}}, \bibinfo {author} {\bibfnamefont
  {A.}~\bibnamefont {Barth{\'{e}}l{\'{e}}my}}, \bibinfo {author} {\bibfnamefont
  {C.}~\bibnamefont {Piamonteze}}, \ and\ \bibinfo {author} {\bibfnamefont
  {M.}~\bibnamefont {Bibes}},\ }\href@noop {} {\bibfield  {journal} {\bibinfo
  {journal} {Phys. Rev. Mater.}\ }\textbf {\bibinfo {volume} {5}},\ \bibinfo
  {pages} {014407} (\bibinfo {year} {2021})}\BibitemShut {NoStop}%
\bibitem [{\citenamefont {Hameed~\textit{et al}.}()}]{HameedSW2021}%
  \BibitemOpen
  \bibfield  {author} {\bibinfo {author} {\bibfnamefont {S.}~\bibnamefont
  {Hameed~\textit{et al}.}},\ }\href@noop {} {\bibinfo  {journal} {{manuscript
  in preparation}}\ }\BibitemShut {NoStop}%
\bibitem [{\citenamefont {Najev}\ \emph {et~al.}(2021)\citenamefont {Najev},
  \citenamefont {Hameed}, \citenamefont {Gautreau}, \citenamefont {Wang},
  \citenamefont {Joe}, \citenamefont {Po\^zek}, \citenamefont {Birol},
  \citenamefont {Fernandes}, \citenamefont {Greven},\ and\ \citenamefont
  {Pelc}}]{Ana2021}%
  \BibitemOpen
\bibfield  {journal} {  }\bibfield  {author} {\bibinfo {author} {\bibfnamefont
  {A.}~\bibnamefont {Najev}}, \bibinfo {author} {\bibfnamefont
  {S.}~\bibnamefont {Hameed}}, \bibinfo {author} {\bibfnamefont {D.~M.}\
  \bibnamefont {Gautreau}}, \bibinfo {author} {\bibfnamefont {Z.}~\bibnamefont
  {Wang}}, \bibinfo {author} {\bibfnamefont {J.}~\bibnamefont {Joe}}, \bibinfo
  {author} {\bibfnamefont {M.}~\bibnamefont {Po\^zek}}, \bibinfo {author}
  {\bibfnamefont {T.}~\bibnamefont {Birol}}, \bibinfo {author} {\bibfnamefont
  {R.~M.}\ \bibnamefont {Fernandes}}, \bibinfo {author} {\bibfnamefont
  {M.}~\bibnamefont {Greven}}, \ and\ \bibinfo {author} {\bibfnamefont
  {D.}~\bibnamefont {Pelc}},\ }\href@noop {} {\bibfield  {journal} {\bibinfo
  {journal} {arXiv:2105.06695}\ } (\bibinfo {year} {2021})}\BibitemShut
  {NoStop}%
\bibitem [{\citenamefont {Taguchi}\ \emph {et~al.}(1993)\citenamefont
  {Taguchi}, \citenamefont {Tokura}, \citenamefont {Arima},\ and\ \citenamefont
  {Inaba}}]{Taguchi1993}%
  \BibitemOpen
  \bibfield  {author} {\bibinfo {author} {\bibfnamefont {Y.}~\bibnamefont
  {Taguchi}}, \bibinfo {author} {\bibfnamefont {Y.}~\bibnamefont {Tokura}},
  \bibinfo {author} {\bibfnamefont {T.}~\bibnamefont {Arima}}, \ and\ \bibinfo
  {author} {\bibfnamefont {F.}~\bibnamefont {Inaba}},\ }\href@noop {}
  {\bibfield  {journal} {\bibinfo  {journal} {Phys. Rev. B}\ }\textbf {\bibinfo
  {volume} {48}},\ \bibinfo {pages} {511} (\bibinfo {year} {1993})}\BibitemShut
  {NoStop}%
\bibitem [{\citenamefont {Yue}\ \emph {et~al.}(2020)\citenamefont {Yue},
  \citenamefont {Quackenbush}, \citenamefont {Laraib}, \citenamefont
  {Carfagno}, \citenamefont {Hameed}, \citenamefont {Prakash}, \citenamefont
  {Thoutam}, \citenamefont {Ablett}, \citenamefont {Lee}, \citenamefont
  {Greven}, \citenamefont {Doty}, \citenamefont {Janotti},\ and\ \citenamefont
  {Jalan}}]{Yue2020}%
  \BibitemOpen
  \bibfield  {author} {\bibinfo {author} {\bibfnamefont {J.}~\bibnamefont
  {Yue}}, \bibinfo {author} {\bibfnamefont {N.~F.}\ \bibnamefont
  {Quackenbush}}, \bibinfo {author} {\bibfnamefont {I.}~\bibnamefont {Laraib}},
  \bibinfo {author} {\bibfnamefont {H.}~\bibnamefont {Carfagno}}, \bibinfo
  {author} {\bibfnamefont {S.}~\bibnamefont {Hameed}}, \bibinfo {author}
  {\bibfnamefont {A.}~\bibnamefont {Prakash}}, \bibinfo {author} {\bibfnamefont
  {L.~R.}\ \bibnamefont {Thoutam}}, \bibinfo {author} {\bibfnamefont {J.~M.}\
  \bibnamefont {Ablett}}, \bibinfo {author} {\bibfnamefont {T.-L.}\
  \bibnamefont {Lee}}, \bibinfo {author} {\bibfnamefont {M.}~\bibnamefont
  {Greven}}, \bibinfo {author} {\bibfnamefont {M.~F.}\ \bibnamefont {Doty}},
  \bibinfo {author} {\bibfnamefont {A.}~\bibnamefont {Janotti}}, \ and\
  \bibinfo {author} {\bibfnamefont {B.}~\bibnamefont {Jalan}},\ }\href@noop {}
  {\bibfield  {journal} {\bibinfo  {journal} {Phys. Rev. Mater.}\ }\textbf
  {\bibinfo {volume} {4}},\ \bibinfo {pages} {112001(R)} (\bibinfo {year}
  {2020})}\BibitemShut {NoStop}%
\bibitem [{\citenamefont {Ulrich}\ \emph {et~al.}(2008)\citenamefont {Ulrich},
  \citenamefont {Ghiringhelli}, \citenamefont {Piazzalunga}, \citenamefont
  {Braicovich}, \citenamefont {Brookes}, \citenamefont {Roth}, \citenamefont
  {Lorenz},\ and\ \citenamefont {Keimer}}]{Ulrich2008}%
  \BibitemOpen
  \bibfield  {author} {\bibinfo {author} {\bibfnamefont {C.}~\bibnamefont
  {Ulrich}}, \bibinfo {author} {\bibfnamefont {G.}~\bibnamefont
  {Ghiringhelli}}, \bibinfo {author} {\bibfnamefont {A.}~\bibnamefont
  {Piazzalunga}}, \bibinfo {author} {\bibfnamefont {L.}~\bibnamefont
  {Braicovich}}, \bibinfo {author} {\bibfnamefont {N.~B.}\ \bibnamefont
  {Brookes}}, \bibinfo {author} {\bibfnamefont {H.}~\bibnamefont {Roth}},
  \bibinfo {author} {\bibfnamefont {T.}~\bibnamefont {Lorenz}}, \ and\ \bibinfo
  {author} {\bibfnamefont {B.}~\bibnamefont {Keimer}},\ }\href@noop {}
  {\bibfield  {journal} {\bibinfo  {journal} {Phys. Rev. B}\ }\textbf {\bibinfo
  {volume} {77}},\ \bibinfo {pages} {113102} (\bibinfo {year}
  {2008})}\BibitemShut {NoStop}%
\bibitem [{\citenamefont {Meijer}\ \emph {et~al.}(1999)\citenamefont {Meijer},
  \citenamefont {Henggeler}, \citenamefont {Brown}, \citenamefont {Becker},
  \citenamefont {Bednorz}, \citenamefont {Rossel},\ and\ \citenamefont
  {Wachter}}]{Meijer1999}%
  \BibitemOpen
  \bibfield  {author} {\bibinfo {author} {\bibfnamefont {G.~I.}\ \bibnamefont
  {Meijer}}, \bibinfo {author} {\bibfnamefont {W.}~\bibnamefont {Henggeler}},
  \bibinfo {author} {\bibfnamefont {J.}~\bibnamefont {Brown}}, \bibinfo
  {author} {\bibfnamefont {O.~S.}\ \bibnamefont {Becker}}, \bibinfo {author}
  {\bibfnamefont {J.~G.}\ \bibnamefont {Bednorz}}, \bibinfo {author}
  {\bibfnamefont {C.}~\bibnamefont {Rossel}}, \ and\ \bibinfo {author}
  {\bibfnamefont {P.}~\bibnamefont {Wachter}},\ }\href@noop {} {\bibfield
  {journal} {\bibinfo  {journal} {Phys. Rev. B}\ }\textbf {\bibinfo {volume}
  {59}},\ \bibinfo {pages} {11832} (\bibinfo {year} {1999})}\BibitemShut
  {NoStop}%
\bibitem [{\citenamefont {Roth}(2008)}]{Roth2008}%
  \BibitemOpen
  \bibfield  {author} {\bibinfo {author} {\bibfnamefont {H.}~\bibnamefont
  {Roth}},\ }\href@noop {} {\bibfield  {journal} {\bibinfo  {journal} {PhD
  Thesis, Universit{\"{a}}t zu K{\"{o}}ln}\ } (\bibinfo {year}
  {2008})}\BibitemShut {NoStop}%
\bibitem [{\citenamefont {de~Vries}\ and\ \citenamefont
  {Wieck}(1995)}]{DeVries1995}%
  \BibitemOpen
  \bibfield  {author} {\bibinfo {author} {\bibfnamefont {D.~K.}\ \bibnamefont
  {de~Vries}}\ and\ \bibinfo {author} {\bibfnamefont {A.~D.}\ \bibnamefont
  {Wieck}},\ }\href@noop {} {\bibfield  {journal} {\bibinfo  {journal} {Am. J.
  Phys.}\ }\textbf {\bibinfo {volume} {63}},\ \bibinfo {pages} {1074} (\bibinfo
  {year} {1995})}\BibitemShut {NoStop}%
\bibitem [{\citenamefont {Hameed}\ \emph {et~al.}(2020)\citenamefont {Hameed},
  \citenamefont {Pelc}, \citenamefont {Anderson}, \citenamefont {Klein},
  \citenamefont {Spieker}, \citenamefont {Yue}, \citenamefont {Das},
  \citenamefont {Ramberger}, \citenamefont {Lukas}, \citenamefont {Liu},
  \citenamefont {Krogstad}, \citenamefont {Osborn}, \citenamefont {Li},
  \citenamefont {Leighton}, \citenamefont {Fernandes},\ and\ \citenamefont
  {Greven}}]{HameedPelc2020}%
  \BibitemOpen
  \bibfield  {author} {\bibinfo {author} {\bibfnamefont {S.}~\bibnamefont
  {Hameed}}, \bibinfo {author} {\bibfnamefont {D.}~\bibnamefont {Pelc}},
  \bibinfo {author} {\bibfnamefont {Z.~W.}\ \bibnamefont {Anderson}}, \bibinfo
  {author} {\bibfnamefont {A.}~\bibnamefont {Klein}}, \bibinfo {author}
  {\bibfnamefont {R.~J.}\ \bibnamefont {Spieker}}, \bibinfo {author}
  {\bibfnamefont {L.}~\bibnamefont {Yue}}, \bibinfo {author} {\bibfnamefont
  {B.}~\bibnamefont {Das}}, \bibinfo {author} {\bibfnamefont {J.}~\bibnamefont
  {Ramberger}}, \bibinfo {author} {\bibfnamefont {M.}~\bibnamefont {Lukas}},
  \bibinfo {author} {\bibfnamefont {Y.}~\bibnamefont {Liu}}, \bibinfo {author}
  {\bibfnamefont {M.~J.}\ \bibnamefont {Krogstad}}, \bibinfo {author}
  {\bibfnamefont {R.}~\bibnamefont {Osborn}}, \bibinfo {author} {\bibfnamefont
  {Y.}~\bibnamefont {Li}}, \bibinfo {author} {\bibfnamefont {C.}~\bibnamefont
  {Leighton}}, \bibinfo {author} {\bibfnamefont {R.~M.}\ \bibnamefont
  {Fernandes}}, \ and\ \bibinfo {author} {\bibfnamefont {M.}~\bibnamefont
  {Greven}},\ }\href@noop {} {\bibfield  {journal} {\bibinfo  {journal}
  {arXiv:2005.00514}\ } (\bibinfo {year} {2020})}\BibitemShut {NoStop}%
\bibitem [{\citenamefont {Wang}\ \emph {et~al.}(2017)\citenamefont {Wang},
  \citenamefont {Huang}, \citenamefont {He}, \citenamefont {Che}, \citenamefont
  {Zhang},\ and\ \citenamefont {Huang}}]{Wang2017}%
  \BibitemOpen
  \bibfield  {author} {\bibinfo {author} {\bibfnamefont {D.}~\bibnamefont
  {Wang}}, \bibinfo {author} {\bibfnamefont {C.}~\bibnamefont {Huang}},
  \bibinfo {author} {\bibfnamefont {J.}~\bibnamefont {He}}, \bibinfo {author}
  {\bibfnamefont {X.}~\bibnamefont {Che}}, \bibinfo {author} {\bibfnamefont
  {H.}~\bibnamefont {Zhang}}, \ and\ \bibinfo {author} {\bibfnamefont
  {F.}~\bibnamefont {Huang}},\ }\href@noop {} {\bibfield  {journal} {\bibinfo
  {journal} {ACS Omega}\ }\textbf {\bibinfo {volume} {2}},\ \bibinfo {pages}
  {1036} (\bibinfo {year} {2017})}\BibitemShut {NoStop}%
\bibitem [{\citenamefont {Zhang}\ \emph {et~al.}(2017)\citenamefont {Zhang},
  \citenamefont {Hao}, \citenamefont {Gao}, \citenamefont {Liu}, \citenamefont
  {Ma}, \citenamefont {Lin}, \citenamefont {Yin},\ and\ \citenamefont
  {Li}}]{Zhang2017}%
  \BibitemOpen
  \bibfield  {author} {\bibinfo {author} {\bibfnamefont {C.}~\bibnamefont
  {Zhang}}, \bibinfo {author} {\bibfnamefont {F.}~\bibnamefont {Hao}}, \bibinfo
  {author} {\bibfnamefont {G.}~\bibnamefont {Gao}}, \bibinfo {author}
  {\bibfnamefont {X.}~\bibnamefont {Liu}}, \bibinfo {author} {\bibfnamefont
  {C.}~\bibnamefont {Ma}}, \bibinfo {author} {\bibfnamefont {Y.}~\bibnamefont
  {Lin}}, \bibinfo {author} {\bibfnamefont {Y.}~\bibnamefont {Yin}}, \ and\
  \bibinfo {author} {\bibfnamefont {X.}~\bibnamefont {Li}},\ }\href@noop {}
  {\bibfield  {journal} {\bibinfo  {journal} {npj Quantum Mater.}\ }\textbf
  {\bibinfo {volume} {2}},\ \bibinfo {pages} {2} (\bibinfo {year}
  {2017})}\BibitemShut {NoStop}%
\end{thebibliography}%

\end{document}